\documentclass[aps,pra,twocolumn,showpacs,floatfix,superscriptaddress]{revtex4-1}
\usepackage{graphicx}
\usepackage{hyperref}
\usepackage{amsmath}
\usepackage{amssymb}
\usepackage{color}
\usepackage{lipsum} 

\begin{document}

 \newcommand{\ket}[1]{|#1\rangle}
 \newcommand{\bra}[1]{\langle #1|}
 
 \newcommand{\ketbra}[2]{\ket{#1}\bra{#2}}
 \newcommand{\biket}[2]{\ket{#1\;#2}}
 \newcommand{\bibra}[2]{\bra{#1\;#2}}
 \newcommand{\biketbra}[2]{\biket {#1}{#2}\bibra {#1}{#2}}
 \newcommand{\preespacio}{\!\text{\hspace{0.2 ex}}}
 \newcommand{\postespacio}{\!\text{\hspace{0.75 ex}}}
 
\newcommand{\app}{a_p^\dagger}
\newcommand{\aap}{a_\mathrm{p}}
\newcommand{\asp}{a_\mathrm{s}^\dagger}
\newcommand{\down}{\ketbra{g}{e}}
\newcommand{\np}{n_\mathrm{p}}
\newcommand{\ns}{n_\mathrm{s}}
\newcommand{\Os}{\Omega_\mathrm{s}}
\newcommand{\gp}{g_\mathrm{p}}
\newcommand{\gs}{g_\mathrm{s}}
\newcommand{\gammap}{\gamma_\mathrm{p}}
\newcommand{\gammas}{\gamma_\mathrm{s}}
\newcommand{\vac}{\mathrm{vac}}
\newcommand{\cnb}[1]{{\color{black} #1}}
\newcommand{\cnbb}[1]{{\color{red} #1}}

\flushbottom
\title{Deterministic Generation of Photon Pairs within the Bad Cavity Limit}

\author{Yue Chang}
\email[]{All the authors contributed equally to the project.}
\affiliation{Max-Planck Institut f\"ur Quantenoptik, Hans-Kopfermann-Str. 1, D-85748 Garching, Germany}
 \author{Alejandro Gonz\'alez-Tudela}
\email[]{All the authors contributed equally to the project.}
\affiliation{Max-Planck Institut f\"ur Quantenoptik, Hans-Kopfermann-Str. 1, D-85748 Garching, Germany}

\author{Carlos S\'{a}nchez-Mu\~{n}oz}
\email[]{All the authors contributed equally to the project.}
\affiliation{Max-Planck Institut f\"ur Quantenoptik, Hans-Kopfermann-Str. 1, D-85748 Garching, Germany}
\affiliation{F\'{\i}sica Te\'{o}rica de la Materia Condensada, Universidad Aut\'{o}noma de Madrid, 28049 Madrid, Spain}
\author{Carlos Navarrete-Benlloch}
\email[]{All the authors contributed equally to the project.}
\affiliation{Max-Planck Institut f\"ur Quantenoptik, Hans-Kopfermann-Str. 1, D-85748 Garching, Germany}
\author{Tao Shi}
\email[]{All the authors contributed equally to the project.}
\affiliation{Max-Planck Institut f\"ur Quantenoptik, Hans-Kopfermann-Str. 1, D-85748 Garching, Germany}

\begin{abstract}
The development, characterization and control of $N$-photon sources are instrumental for quantum technological applications. This work constitutes a step forward in this direction, \cnb{where we propose a cavity quantum electrodynamics setup designed for the generation of photon pairs. We analyze it both via the scattering and master equation formalisms. From the connection between these two frameworks it naturally arises a physical criterion characterizing when weakly-driven systems behave as continuous antibunched two-photon sources. We find the optimal parameters for which our setup works as an efficient photon-pair source, showing also that it becomes a deterministic down-converter of single photons. We provide a specific implementation based on state-of-the-art superconducting circuits, showing how our proposal is within the reach of current technologies.}
\end{abstract}

\maketitle

\textbf{Introduction.} 
Single-photon sources \cite{lounis05a} are one of the cornerstones of many quantum information protocols \cite{obrien07a,obrien09a}. The success in the fabrication of these sources is built upon the simple non-linear systems required, e.g., a two-level system \cite{brunel99a}, and a well-established characterization through the well-known second-order coherence function \cite{glauber63a} $g^{(2)}(\tau)$ (see the explicit definition below), which yields $g^{(2)}(0)=0$ for a perfect single photon source. The extension to $N$-photon sources, especially $N=2$, lies also at the heart of many recent quantum technological applications such as the generation of NOON states \cite{afek10a} instrumental for quantum metrology \cite{giovannetti04a}, beating the diffraction limit~\cite{dangelo01a}, or even biological purposes \cite{denk90a,horton13a}. There exist several methods to generate multiphoton states, for example, probabilistic schemes using down-converted photons \cite{bruno14a} and postselection to build up 
higher photon numbers \cite{ourjoumtsev06b,zavatta08a, yao12a}, but typically at the price of an small probability. An alternative consists in using atom-like systems strongly coupled to cavities \cite{law96a} or biexciton states in quantum dots systems \cite{callsen13a,dousse10a,muller14a}. The former has shown spectacular advances in the microwave regime for intracavity fields \cite{hofheinz08a,hofheinz09a}, but whose extension to traveling photons is so far limited to single photons \cite{johnson10a,bozyigit11a,eichler11a,pechal14a}. Therefore, the efficient generation of multiphoton states is still an open question which attracts a lot of attention \cite{sanchezmunoz14a,sanchezmunoz15a,gonzaleztudela15d}.

Moreover, the characterization of multiphoton sources is done either via full-state tomography \cite{ourjoumtsev06b,zavatta08a,hofheinz09a,eichler11a,pechal14a}, experimentally very demanding, or through standard correlations functions \cite{rundquist14a,koch11a,bozyigit11a} such as $g^{(n)}(\tau)$. \cnb{However, as recently shown in \cite{sanchezmunoz14a} in the context of the driven Jaynes-Cummings ladder in the strong-coupling regime, such standard correlations fail to capture unambiguously the multiphoton character of its emission (see also \cite{sanchezmunoz15a}). The authors define then two quantities that indeed accomplish this task in their system. First, the \textit{purity} of $N$-photon emission, which represents the percentage of the emission that exists grouped in $N$-photon bundles. The statistics of the emission was then shown to be captured by a generalized second-order coherence function for the $N$-photon bundles, $g^{(2)}_N(\tau)$ (see below for a definition in the $N=2$ case), which in the system studied by the authors revealed a crossover between Poisson and antibunched statistics (that is, between an $N$-photon \textit{laser} and a \textit{gun}) as the lifetime of the two-level system is increased.} Despite this progress, a unified and formal characterization of these sources is lacking and one still finds different definitions in the literature \cite{hong10a,koch11a,rundquist14a,leboite14a,sanchezmunoz14a}.

\begin{figure}[tb]
\includegraphics[width=0.85\linewidth]{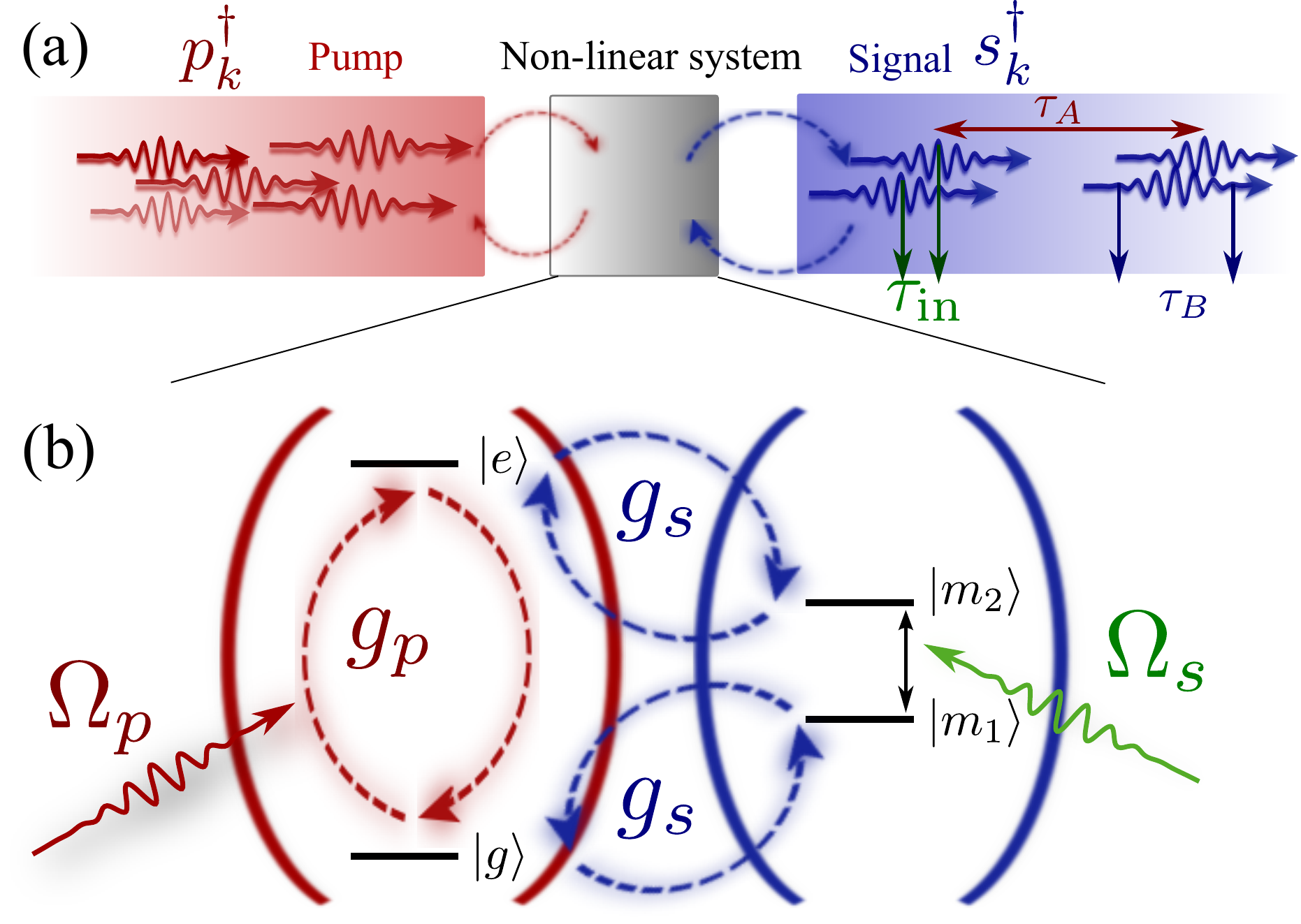}
\caption{Scheme for photon-pair generation. (a) The nonlinear system is driven through the pump bath (left, red) and the emission of photons coming out through the signal bath (right, blue) is analyzed. We depict the three relevant timescales that characterize the emission in our system: the intrinsic width of the single-photon wavepackets ($\tau_B$), and the separation between the pairs ($\tau_A$) and between the photons within the same pair ($\tau_{\mathrm{in}}$). (b) Scheme of the nonlinear system that provides the interface between incoming and outgoing photons.}
\label{fig:1}
\end{figure}

\cnb{In this Letter, we introduce a cavity quantum electrodynamics setup that acts as a continuous source of photon pairs when weakly driven (and as a deterministic down-converter when excited by single photons), and does so within the \textit{bad-cavity limit}. We propose a general practical criterion that characterizes when such weakly-driven systems behave as emitters of photon pairs in well-defined pulses. Our analysis is based on the connection between two different perspectives, the scattering and master equation formalisms  \cite{shi09a,caneva15a,shi15a}, whose link allows us to show how the two- and four-photon scattering wavefunctions are equivalent, respectively, to the unnormalized correlation functions $G^{(2)}(\tau)$ and $G^{(2)}_2(\tau)$ of weakly-driven continuous sources. We analyze the scalings and figures of merit of the proposal, and finally discuss possible implementations focusing on currently available circuit QED architectures.}

\textbf{Criterion for continuous single photon-pair source.} 
We consider the general scheme for a source depicted in Fig.~\ref{fig:1}(a). A non-linear system $S$ is coupled to two (one-dimensional) baths \cite{shi15a}. The \textit{pump} bath is used to excite the system with, e.g., a continuous driving or pulses with well defined photon number, while the emitted light is monitored through the \textit{signal} bath.
Working in a picture rotating at some characteristic frequency of the system $k_p$ that we will choose later (we take $\hbar=c=1$ so that frequency and momentum, or time and space, can be used indistinctively), and denoting by $\{p_k,s_k\}_{k\in\mathbb{R}}$ the annihilation operators of the baths, the Hamiltonian is given by $H=H_S+H_B+H_{S\hspace{-0.2mm}B}$, with Hamiltonians $H_S$ and $H_B=\int\hspace{-1mm}dk \hspace{0.8mm} k (p_k^\dagger p_k + s_k^\dagger s_k)$ for system and bath, respectively, which interact through
\begin{equation}
H_{S\hspace{-0.2mm}B}=\int\hspace{-1.5mm}dk \left(\sqrt{\frac{\gamma_p}{2 \pi}}p_k^\dagger a_p+\sqrt{\frac{\gamma_s}{2 \pi}}s_k^\dagger a_s\right)+\mathrm{H.c.}\,,
\end{equation}
where $a_j$ is the system operator that couples to the signal/pump ($j=s/p$) bath.

The scattering formalism \cite{sakurai14a} is naturally suited for analyzing processes such as the conversion of a given input state $|k_1,...,k_m\rangle_p=p_{k_1}^\dagger...p_{k_m}^\dagger|0\rangle$ with $m$ incoming pump photons with momenta $\{k_1,...,k_m\}$, into an outgoing state $|q_1,...,q_n\rangle_s=s_{q_1}^\dagger...s_{q_n}^\dagger|0\rangle$ with $n$ signal photons with momenta $\{q_1,...,q_n\}$. All the asymptotic information is contained in the so-called $S$-matrix, defined as $S=\lim\hspace{0.05mm}_{t_i\rightarrow-\infty}^{t_f\rightarrow+\infty}e^{iH_Bt_f}e^{-iH(t_f-t_i)}e^{-iH_Bt_i}$. For example, the probability amplitude of the previous process is given by $_s\langle q_1,...,q_n|S|k_1,...,k_m\rangle_p$ as explained in the supplemental material \cite{supmat}.

An alternative scenario is that in which the system is continuously driven by a monochromatic laser at some frequency $k_0$ through the pump bath (in the picture rotating with $k_p$). A standard approach in this context consists on integrating the bath degrees of freedom, what results in an master equation for the system's state $\rho$ \cite{gardiner_book00a,carmichael_book02a}:
\begin{equation}
\dot{\rho}=-i\hspace{-0.3mm}\left[H_S\hspace{-0.8mm}+\hspace{-1mm}H_D,\rho\right]+\hspace{-0.8mm}\sum_{j=s,p}\hspace{-1mm} \frac{\gamma_j}{2} (2 a_j\rho a^\dagger_j-a^\dagger_j a_j\rho-\rho a^\dagger_j a_j)\,,
\label{eq:master-eq1}
\end{equation}
where $H_D=\Omega_p\left(e^{-i k_0 t}a_p^\dagger+\mathrm{H.c.}\right)$ is an extra driving term, being $\Omega_p$ its amplitude (taken real and positive without loss of generality). The statistics of the field emitted through the baths can be analyzed through multi-time correlation functions which, using input-output theory \cite{gardiner_book00a,carmichael_book02a}, can be ultimately related to system correlators $G^{(n)}_j(\tau_1,...,\tau_n)=\langle a_j^\dagger(\tau_1)...a_j^\dagger(\tau_{n}) a_j(\tau_n)...a_j(\tau_1)\rangle$, where $\tau_1<...<\tau_n$ with the operators defined in the Heisenberg picture. The master equation allows evaluating these functions via the quantum regression theorem \cite{gardiner_book00a,carmichael_book02a}.

Even though these two formalisms (scattering and master equation) seem to apply to very different scenarios, they are very much connected \cite{caneva15a,shi15a}. For example, let us consider a situation in which $H$ connects pump photons with signal photon-pairs, as is the case of the system that we introduce in the next section. Using the various definitions provided above, we have been able to find a relation between scattering amplitudes and correlation functions of the system under weak driving \cite{supmat}. In the case of the second-order correlation function, to first nontrivial order in $\Omega_p$ we get \cite{supmat}
\begin{equation}
G_s^{(2)}(\tau)=\lim_{t\rightarrow\infty}G_s^{(2)}(t,t+\tau)\propto\Omega_p^2| \Psi_{\mathrm{2ph}}(x_1,x_2)|^2\,,
\end{equation}
where $x_1-x_2=\tau$, and we introduce the two-photon wave function $\Psi_{\mathrm{2ph}}(x_1,x_2)=\langle 0|s(x_1)s(x_2)S|k_0\rangle_p$, with $s(x)=(2\pi)^{-1/2}\int\hspace{-1mm}dk\hspace{0.7mm}s_k e^{ikx}$ annihilating signal excitations in real space. We find a similar connection between the photon-pair second-order correlation function~\cite{sanchezmunoz14a} and the four-photon wavefunction $\Psi_{\mathrm{4ph}}(x_1,x_2,x_3,x_4)=\langle 0|s(x_1)s(x_2)s(x_3)s(x_4)S|k_0,k_0\rangle_p$. This reads \cite{supmat}
\begin{eqnarray}
G_{s,2}^{(2)}(\tau)=&\lim_{t\rightarrow\infty}&G_s^{(4)}(t,t,t+\tau,t+\tau)
\\
&&\propto \Omega_p^4|\Psi_{\mathrm{4ph}}(x_1,x_1,x_2,x_2)|^2\,, \nonumber
\end{eqnarray}
with $x_1-x_2=\tau$. 

Within the scattering formalism, it is clear that the system will be an ideal single photon-pair source whenever $\vert\Psi_\mathrm{2ph}(x_1,x_2)\vert$ and $\vert\Psi_\mathrm{4ph}(x_1,x_1,x_2,x_2)\vert$ are, respectively, single-narrowly-peaked and vanishing at $x_1=x_2$, meaning that the system scatters photons in well-spaced wave packets containing two signal photons. Therefore, the equivalences above \cnb{naturally give rise to} a physical criterion characterizing when the weakly-driven system is emitting light in photon pairs: $G_s^{(2)}(0)>G_s^{(2)}(\tau)$, while keeping $G_{s,2}^{(2)}(0)<G_{s,2}^{(2)}(\tau)$, that is, it has to show bunching of single photons, but anti-bunching between photon pairs. Moreover, the antibunching timescale must be larger than the bunching one, so that separation between the pairs is guaranteed. This connection provides then formal grounds to the \cnb{use of the photon-pair second-order correlation function \cite{sanchezmunoz14a} in weakly-driven systems}.

\begin{figure*}[tb]
\begin{center}
\includegraphics[width=\linewidth]{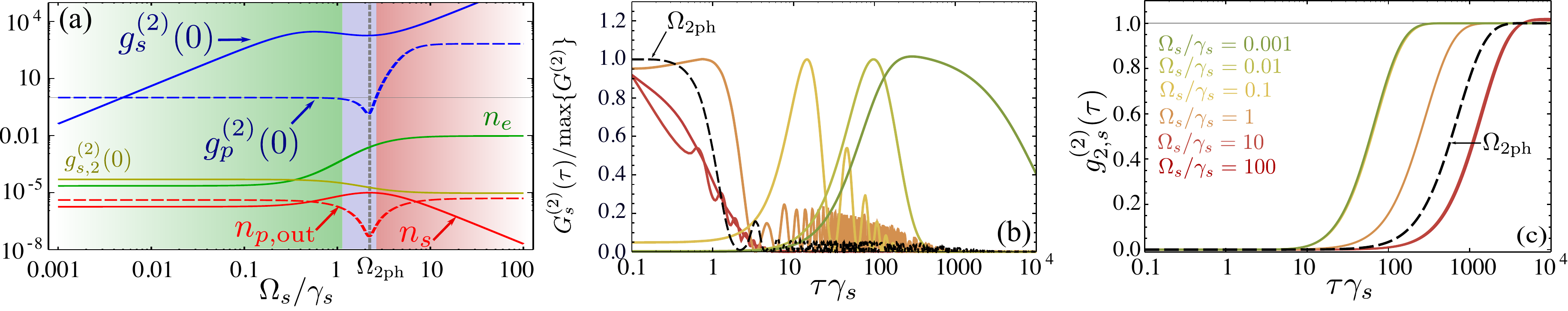}
\end{center}
\caption{(a) Main steady-state observables as a function of the control drive $\Omega_s$. Parameters are $\gamma_p=20\gamma_s$, $\Omega_p=0.01\gamma_s$, $\gamma^*=0$, $g_p=g_\mathrm{s}=0.1 \gamma_s$. (b) and (c) show, respectively, $G^{(2)}_{s}(\tau)$ normalized to its maximal value and $g^{(2)}_{s,2}(\tau)$ as a function of $\tau$ for $\Omega_s$ changing logarithmically from $0.001\gammas$ to $100\gammas$ (in color) and for the optimal condition $\Omega_{\mathrm{2ph}}$ (dashed black) defined in Eq.~\ref{cond}. Notice that in (c) the curves from $\Omega_s=10^{-3}$ to $0.1$ and $10$ to $100$ overlap.}
\label{fig:2}
\end{figure*}

\textbf{Engineered photon-pair source by quantum interference.}
Let us now introduce a cavity QED model designed to emit photon pairs in well separated wavepackets, of which we provide a specific implementation below. Consider the nonlinear system of Fig.~\ref{fig:1}(b), consisting of a four-level system with states $\{\ket{g},\ket{m_1},\ket{m_2},\ket{e}\}$, coupled to two photonic modes in independent cavities and a classical field. The \textit{pump mode}, with annihilation operator $a_p$, is resonant with the $\ket{g}\hspace{-1mm}\leftrightarrow\hspace{-1mm}\ket{e}$ transition. The \textit{signal mode}, on the other hand, has annihilation operator $a_s$ and is resonant both with $\ket{g}\hspace{-1mm}\leftrightarrow\hspace{-1mm}\ket{m_1}$ and $\ket{m_2}\hspace{-1mm}\leftrightarrow\hspace{-1mm}\ket{e}$. Finally, the classical field controls resonantly the transition $\ket{m_1}\hspace{-1mm}\leftrightarrow\hspace{-1mm}\ket{m_2}$ with a Rabi frequency $\Omega_s$ (taken positive 
for simplicity),
 and will 
allow us to tune between different regimes of emission, specifically to optimize the two-photon emission.

In a picture rotating at the pump frequency, the system is then described by the Hamiltonian:
\begin{eqnarray}
H_S &=& g_p\, a_p^\dagger \ketbra{g}{e} + \Omega_s\ketbra{m_2}{m_1} \nonumber
\\
&&\hspace{0.3cm}+g_s\, a_s^\dagger \left(\ketbra{m_2}{e}+\ketbra{g}{m_1}\right)+\mathrm{H.c}\,.
\label{eq:hamiltonian}
\end{eqnarray}
The advantage of using a bimodal configuration is twofold: i) it allows us to separate the pump/signal fields, which we assume to couple to different baths; and ii) it will allow the system to act as a good antibunched two-photon source within the \emph{bad-cavity limit}, that is, $g_j\ll \gamma_j$, but still with cooperativities $C_j=g_j^2/\gamma_j\gamma^*> 1$, where $\gamma^*$ is the spontaneous emission rate of the four-level system to other dissipative channels. This is in contrast to previous proposals relying on the more demanding \emph{strong-coupling} regime $g_j>\gamma_j$.

Let us first analyze the ideal regime with $\gamma^*=0$. In Fig. \ref{fig:2}(a) we show the dependence of the main steady state observables on the control drive $\Omega_s$, as obtained from the master equation (\ref{eq:master-eq1}) or its connection with scattering theory \cite{supmat}, and using representative parameters within the bad-cavity limit. We represent various populations $n_j$ ($j=p$ for pump, $s$ for signal, and $e$ for excited state), including the one for the output pump mode $a_{p,\mathrm{out}}=2\Omega_p/\gamma_p-i a_p$, as well as normalized correlation functions $g_j^{(2)}(\tau)=G_j^{(2)}(\tau)/n_j^2$ and $g_{s,2}^{(2)}(\tau)=G_{s,2}^{(2)}(\tau)/[G^{(2)}_s(0)]^2$ at $\tau=0$.  We can differentiate three regimes identified through the second-order correlation function of the pump. i) $g_{p}^{(2)}(0)=1$, green background: This region shows a transition from $g_{s}^{(2)}(0)<1$, where the signal cavity is therefore emitting single photons, to $g_{s}^{(2)}(0)>1$, which corresponds to correlated emission 
from the cascade through the intermediate levels. However, when looking at the dynamics of $G_s^{(2)}(\tau)$ in Fig. \ref{fig:2}(b), we can see how the maximum two-photon probability occurs at $\tau>0$, and therefore it is still not a good photon-pair source, since this indicates that the photons inside a pair are spatially separated. ii) $g_{p}^{(2)}(0)<1$, blue background: This region shows $g_{s,2}^{(2)}(0)<g_{s,2}^{(2)}(\tau)$, maximal $G_{s}^{(2)}(\tau)$  very close to $\tau=0$ and a bunching timescale much shorter than the antibunching one of $g_{s,2}^{(2)}(\tau)$. Therefore, photons inside a pair are emitted together and the pairs are well separated, so the system behaves as a good photon-pair source according to the criterion defined above. Moreover, this region features a maximal signal population $n_s$ (and minimal $n_{p,\mathrm{out}}$) at the optimum control drive $\Omega_{\mathrm{2ph}}$ (that we specify below), yielding then a 
maximum photon-pair emission rate given by $\gamma_s n_s$. Remarkably, this condition also leads to the deterministic down-conversion of single-photon pulses, as we will show later. iii) $g_{p}^{(2)}(0)>1$, red background: With photon-pair emission but with a decrease of its rate.

In order to gain understanding on the two-photon emission process, we analyze the relevant timescales for the emission of photon pairs, which are schematically depicted in Fig. \ref{fig:1}(a) and we define in what follows. Once the system arrives at $\ket{e}$ it relaxes to $\ket{g}$ in a time $\tau_A$ (acting then as the \textit{reloading} time), either emitting a pump photon or two cascaded signal photons separated by a time $\tau_\mathrm{in}$. Denoting by $\tau_B$ the intrinsic width of the single-photon wavepackets, it is then clear that $\tau_{\mathrm{in}}<\tau_B<\tau_A$ is required for the system to act as an antibunched two-photon source. We have made a detailed analysis of these timescales \cite{supmat}, and here we focus on the results found within the bad-cavity limit ($g_j\ll\gamma_j$) and with a strong-enough control drive ($\Omega_s\gg g_j^2/\gamma_j$). In this regime, it can be shown that $\tau_\mathrm{in}$ scales proportional to $\Omega_s^{-1}$, and hence the delay between photons within the 
same pair can be made 
arbitrarily small by increasing the control drive. Under such circumstances, we get $\tau_B\sim\gamma_s^{-1}$, which determines the timescale of the bunching in $G_s^{(2)}(\tau)$, and provides the requirement $\Omega_s>\gamma_s$. Finally, $\tau_A$ is determined by the decay rate from $\ket{e}$ to $\ket{g}$, which can occur through two different paths: either mediated by the pump cavity at the corresponding Purcell-enhanced rate $\Gamma_p=4g_p^2/\gamma_p$, or through the signal cavity at a rate $\Gamma_s(\Omega_s)=4g_s^2\gamma_s/(\gamma_s^2+4\Omega_s^2)$, that we easily estimate by adiabatically eliminating the signal/pump mode and the intermediate levels \cite{supmat} (which are where both weakly populated). We obtain then $\tau_A^{-1}\approx\Gamma_p+\Gamma_s(\Omega_s)$. Notice that such an adiabatic elimination turns the system into an effective coherently-driven two-level system, defined by $\{\ket{g},\ket{e}\}$, with two decay paths into the $p/s$ channels with decay rates $\Gamma_p$ and $\Gamma_s(\Omega_
s)$. 
Using this simplified model, we can show that optimal control driving $\Omega_\mathrm{2ph}$ leading to the minimum $g_{p}^{(2)}(0)$ (and $n_{p,\mathrm{out}}$) is determined by a quantum interference effect occurring when the rates of both decay paths are equal, that is, $\Gamma_s(\Omega_{\mathrm{2ph}})=\Gamma_p$, condition that provides an optimal control drive
\begin{align}
\Omega_\mathrm{2ph}^2\approx\gamma_s^2[\Gamma_s(0)-\Gamma_p]/4\Gamma_p\,,
\label{cond}
\end{align}
where we  see that $\Gamma_p<\Gamma_s(0)= 4g_s^2/\gamma_s$ is required. This is a similar interference effect as the one used in previous works to enhance single-photon nonlinearities \cite{liew10a,bamba11a,majumdar12a,shi13a}.

\begin{figure}[tb]
\begin{center}
\includegraphics[width=0.85\columnwidth]{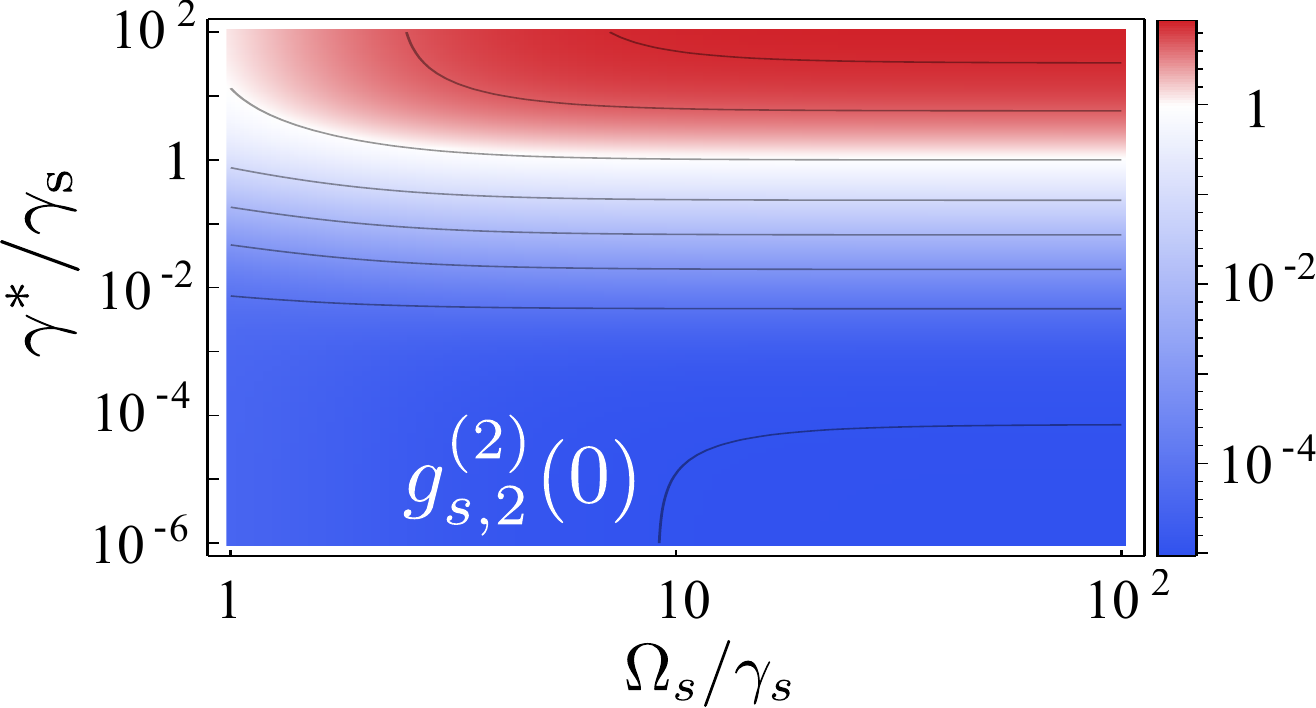}
\end{center}
\caption{Density plot of the second-order correlation of the pairs at zero delay  $g^{(2)}_{2,s}(0)$ as a function of $\Omega_s$ and $\gamma^*$, with the rest of parameters fixed as in Fig. \ref{fig:2}.}
\label{fig:3}
\end{figure}
Remarkably, the optimal control driving $\Omega_\mathrm{2ph}$ can be also found within the scattering formalism \cite{supmat} from the condition $_p\langle k_f|S|k_i\hspace{-1mm}=\hspace{-1mm}0\rangle_p\hspace{-1mm}=\hspace{-1mm}0\hspace{2mm}\forall k_f$, that is, by demanding an incoming resonant pump photon to be transformed into two signal photons with probability 1 (\textit{deterministic down-conversion}). Moreover, it is possible \cite{supmat} to show that the \textit{reflection coefficient} $\vert\hspace{-1mm}\int\hspace{-1mm}dk\hspace{0.5mm}_p\hspace{-0.5mm}\langle k|S|k_i\rangle_{\hspace{-0.4mm}p}\vert^2$ has a Lorentzian shape as a function of the incident momentum $k_i$, with a width $\sim \Gamma_p+\Gamma_s(\Omega_s)$, that provides the bandwidth for efficient down-conversion of single-photon pulses.

 \textbf{Finite cooperativity.} 
To further estimate the feasibility of our proposal, we now analyze the effect of having spontaneous emission from $\ket{e}$ to $\ket{g}$ through other dissipative channels not considered in the ideal model. Assuming such processes to occur at rate $\gamma^*\ll\gamma_{p,s}$, this time-scale contributes to the reloading time as $\tau_A^{-1}\approx\Gamma_s(\Omega_s)+\Gamma_p+\gamma^*$ \cite{supmat}. It is then clear that as long as $\gamma^*\ll\gamma_s$ the condition $\tau_A\ll\{\tau_B,\tau_\mathrm{in}\}$ will then still be satisfied. Such an intuitive result is confirmed by the simulation of Fig. \ref{fig:3}, where we show $g_{s,2}^{(2)}(0)$ as a function of the control driving $\Omega_s$ and $\gamma^*$, as obtained directly from numerical simulation of the master equation. As expected, by increasing $\gamma^*$ above $\gamma_s$ the system shows a transition from antibunched to bunched photon pairs, as $\tau_A$ becomes comparable to $\tau_B$.

\textbf{Implementation.} 
\begin{figure}[tb]
\begin{center}
\includegraphics[width=0.99\columnwidth]{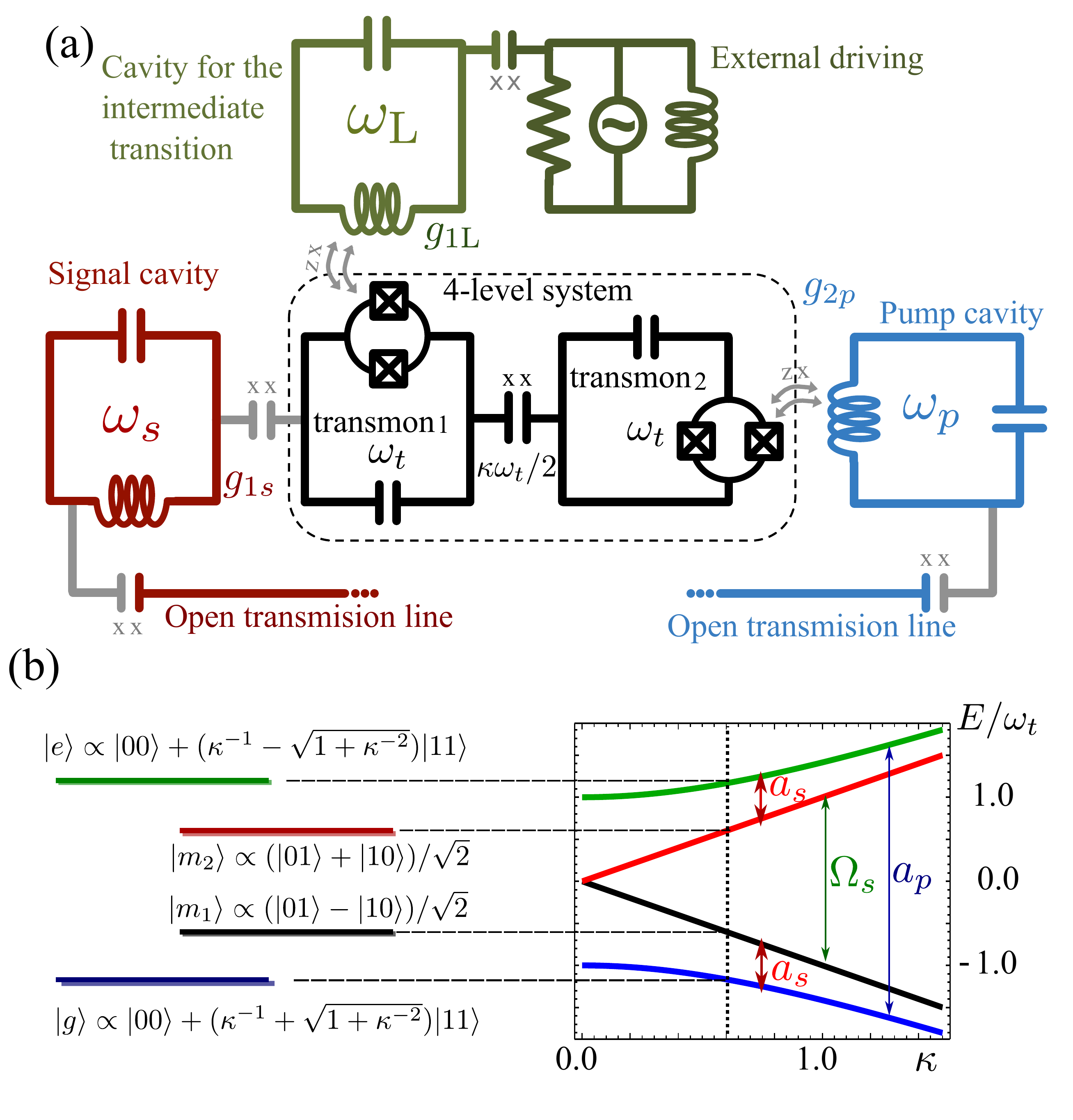}
\end{center}
\caption{(a) Circuit QED implementation: two coupled transmon qubits provide the desired four-level structure (b), while three LC resonators provide the single-mode cavities playing the role of signal, pump, and the classical driving for the intermediate transition. The baths are implemented with open transmission lines.}
\label{fig:4}
\end{figure}
One versatile platform to implement our ideas is circuit QED \cite{xiang13a,devoret13a,schoelkopf08a}, where we can take advantage of long-lived qubits, single-mode cavities, and open transmissions lines to design our proposed setup. A concrete implementation is depicted in Fig. \ref{fig:4}(a). Two identical qubits (transmons \cite{koch07a,schreier08a} in the figure) with energies $\omega_t$ are capacitively coupled through an $xx$ interaction $\omega_t\kappa\sigma_x^{(1)}\sigma_x^{(2)}/2$, whose spectrum is shown in Fig. \ref{fig:4}(b). We see that the desired four level structure appears between the states $\ket{m_{2,1}}\propto\ket{10}\pm\ket{01}$ and $\ket{e,g}\propto (\kappa^{-1}\mp\sqrt{1+\kappa^{-2}})\ket{11}+\ket{00}$, with respective energies $E_{2,1}=\pm\omega_t\kappa$ and $E_{e,g}=\pm\omega_t\sqrt{1+\kappa^2}$. Two single-mode LC resonators with frequencies $\omega_p=E_e-E_g$ and $\omega_s=E_e-E_2=E_1-E_g$ provide the pump and signal modes, respectively, while an additional strongly-driven 
resonator with frequency $\omega_\mathrm{L}=E_2-E_1$ is used to control the intermediate transition. The pump and classical cavities are inductively coupled each to one transmon via an $xz$ interaction, $g_{2p}(a_p+a_p^\dagger)\sigma_z^{(2)}$ and $g_{1\mathrm{L}}(a_\mathrm{L}+a_\mathrm{L}^\dagger)\sigma_z^{(1)}$. Finally, the signal cavity is capacitively coupled through an $xx$ interaction to one of the transmons, $g_{1s}(a_s+a_s^\dagger)\sigma_x^{(1)}$. Working in the $\kappa\ll 2$ regime and provided $\{g_{1\mathrm{L}}\alpha,g_{2p}\kappa,g_{1s}/\sqrt{2}\}\ll 2\omega_t\kappa$, these type of couplings ensure that in the eigenbasis of the coupled-qubit system the full system Hamiltonian takes the form of Eq. (\ref{eq:hamiltonian}), with $\Omega_s=g_{1\mathrm{L}}\alpha$, $g_p=-g_{2p}\kappa$, and $g_s=g_{1s}/\sqrt{2}$ \cite{supmat}, being $\alpha$ the number of excitations in the classical cavity that can be controlled via the external driving.
Using $\gamma_{p,s}$ on the tens of MHz range, the spontaneous emission of the superconducting qubits gets orders of magnitude below the large cavity decay rates, while the cooperativities can be made very large since couplings $g_{2p}$ and $g_{1s}$ up to tens of MHz are routinely achieved in current experiments. With these parameters, at the optimal point $\Omega_s=\Omega_\mathrm{2ph}$ the rate of photon-pair emission $\gamma_s n_s$ can get up to the 0.1-10KHz range for $\Omega_p/\gamma_s \in [10^{-2},10^{-1}]$.

Other platforms may fulfill our requisites in the optical domain, such as natural or artificial atoms using its ``butterfly''-like level structure, coupled to nanophotonic cavities. Current experiments with atoms  \cite{thompson13a} show cooperativities $C\approx 10$ with decay rates up to $25$GHz, which would lead to photon-pair emission rates in the 0.1-10MHz range, which exceed current parametric down-conversion technologies. Remarkably, the expected bandwidth matches the characteristic atomic linewidths, usually laying in the MHz range.
 
\textbf{Conclusions.} \cnb{By exploiting the connection between scattering theory and the master equation we have formalized a criterion characterizing photon-pair sources under weak driving. The criterion is based on the dynamics of the standard correlation function $G^{(2)}(\tau)$ and the generalized one of the pairs, $G^{(2)}_2(\tau)$, first introduced in \cite{sanchezmunoz14a}.} Moreover, we have provided a cavity QED setup that acts as a deterministic down-converter when excited by single photons or as a continuous photon-pair source when weakly driven, and does so within the bad-cavity limit. Our analysis of the figures of merit and scaling with different parameters has shown the feasibility of the proposal, for which we have designed a concrete implementation based on superconducting circuits. We believe that our characterization, analysis, and implementation proposal represent an important step forward in the fabrication of efficient two-photon sources, which can in addition be extended to higher photon numbers.

\textbf{Acknowledgements}. The authors thank Peter Zoller and Eugenio Rold\'an for useful suggestions. The work of Y.C., A.G.-T., C.N.-B. and T.S. was funded by the European Union integrated project \emph{Simulators and Interfaces with Quantum Systems} (SIQS). A.G.-T. also acknowledges support from Intra-European Marie-Curie Fellowship NanoQuIS (625955). C. S. M. is supported by the FPI programme of the Spanish MINECO through projects MAT2011- 22997 and MAT2014-53119-C2-1-R.

\bibliography{Sci,books,arXiv}

\begin{thebibliography}{51}%
\makeatletter
\providecommand \@ifxundefined [1]{%
 \@ifx{#1\undefined}
}%
\providecommand \@ifnum [1]{%
 \ifnum #1\expandafter \@firstoftwo
 \else \expandafter \@secondoftwo
 \fi
}%
\providecommand \@ifx [1]{%
 \ifx #1\expandafter \@firstoftwo
 \else \expandafter \@secondoftwo
 \fi
}%
\providecommand \natexlab [1]{#1}%
\providecommand \enquote  [1]{``#1''}%
\providecommand \bibnamefont  [1]{#1}%
\providecommand \bibfnamefont [1]{#1}%
\providecommand \citenamefont [1]{#1}%
\providecommand \href@noop [0]{\@secondoftwo}%
\providecommand \href [0]{\begingroup \@sanitize@url \@href}%
\providecommand \@href[1]{\@@startlink{#1}\@@href}%
\providecommand \@@href[1]{\endgroup#1\@@endlink}%
\providecommand \@sanitize@url [0]{\catcode `\\12\catcode `\$12\catcode
  `\&12\catcode `\#12\catcode `\^12\catcode `\_12\catcode `\%12\relax}%
\providecommand \@@startlink[1]{}%
\providecommand \@@endlink[0]{}%
\providecommand \url  [0]{\begingroup\@sanitize@url \@url }%
\providecommand \@url [1]{\endgroup\@href {#1}{\urlprefix }}%
\providecommand \urlprefix  [0]{URL }%
\providecommand \Eprint [0]{\href }%
\providecommand \doibase [0]{http://dx.doi.org/}%
\providecommand \selectlanguage [0]{\@gobble}%
\providecommand \bibinfo  [0]{\@secondoftwo}%
\providecommand \bibfield  [0]{\@secondoftwo}%
\providecommand \translation [1]{[#1]}%
\providecommand \BibitemOpen [0]{}%
\providecommand \bibitemStop [0]{}%
\providecommand \bibitemNoStop [0]{.\EOS\space}%
\providecommand \EOS [0]{\spacefactor3000\relax}%
\providecommand \BibitemShut  [1]{\csname bibitem#1\endcsname}%
\let\auto@bib@innerbib\@empty
\bibitem [{\citenamefont {Lounis}\ and\ \citenamefont
  {Orrit}(2005)}]{lounis05a}%
  \BibitemOpen
  \bibfield  {author} {\bibinfo {author} {\bibfnamefont {B.}~\bibnamefont
  {Lounis}}\ and\ \bibinfo {author} {\bibfnamefont {M.}~\bibnamefont {Orrit}},\
  }\href@noop {} {\bibfield  {journal} {\bibinfo  {journal} {Reports on
  Progress in Physics}\ }\textbf {\bibinfo {volume} {68}},\ \bibinfo {pages}
  {1129} (\bibinfo {year} {2005})}\BibitemShut {NoStop}%
\bibitem [{\citenamefont {O'Brien}(2007)}]{obrien07a}%
  \BibitemOpen
  \bibfield  {author} {\bibinfo {author} {\bibfnamefont {J.~L.}\ \bibnamefont
  {O'Brien}},\ }\href@noop {} {\bibfield  {journal} {\bibinfo  {journal}
  {Science}\ }\textbf {\bibinfo {volume} {318}},\ \bibinfo {pages} {1567}
  (\bibinfo {year} {2007})}\BibitemShut {NoStop}%
\bibitem [{\citenamefont {O'Brien}\ and\ \citenamefont {{Akira
  Furusawa}}(2009)}]{obrien09a}%
  \BibitemOpen
  \bibfield  {author} {\bibinfo {author} {\bibfnamefont {J.~L.}\ \bibnamefont
  {O'Brien}}\ and\ \bibinfo {author} {\bibfnamefont {J.~V.}\ \bibnamefont
  {{Akira Furusawa}}},\ }\href@noop {} {\bibfield  {journal} {\bibinfo
  {journal} {Nature Photonics}\ }\textbf {\bibinfo {volume} {3}},\ \bibinfo
  {pages} {687} (\bibinfo {year} {2009})}\BibitemShut {NoStop}%
\bibitem [{\citenamefont {Brunel}\ \emph {et~al.}(1999)\citenamefont {Brunel},
  \citenamefont {Lounis}, \citenamefont {Tamarat},\ and\ \citenamefont
  {Orrit}}]{brunel99a}%
  \BibitemOpen
  \bibfield  {author} {\bibinfo {author} {\bibfnamefont {C.}~\bibnamefont
  {Brunel}}, \bibinfo {author} {\bibfnamefont {B.}~\bibnamefont {Lounis}},
  \bibinfo {author} {\bibfnamefont {P.}~\bibnamefont {Tamarat}}, \ and\
  \bibinfo {author} {\bibfnamefont {M.}~\bibnamefont {Orrit}},\ }\href@noop {}
  {\bibfield  {journal} {\bibinfo  {journal} {Physical Review Letters}\
  }\textbf {\bibinfo {volume} {83}},\ \bibinfo {pages} {2722} (\bibinfo {year}
  {1999})}\BibitemShut {NoStop}%
\bibitem [{\citenamefont {Glauber}(1963)}]{glauber63a}%
  \BibitemOpen
  \bibfield  {author} {\bibinfo {author} {\bibfnamefont {R.~J.}\ \bibnamefont
  {Glauber}},\ }\href@noop {} {\bibfield  {journal} {\bibinfo  {journal}
  {{Phys. Rev. Lett.}}\ }\textbf {\bibinfo {volume} {10}},\ \bibinfo {pages}
  {84} (\bibinfo {year} {1963})}\BibitemShut {NoStop}%
\bibitem [{\citenamefont {Afek}\ \emph {et~al.}(2010)\citenamefont {Afek},
  \citenamefont {Ambar},\ and\ \citenamefont {Silberberg}}]{afek10a}%
  \BibitemOpen
  \bibfield  {author} {\bibinfo {author} {\bibfnamefont {I.}~\bibnamefont
  {Afek}}, \bibinfo {author} {\bibfnamefont {O.}~\bibnamefont {Ambar}}, \ and\
  \bibinfo {author} {\bibfnamefont {Y.}~\bibnamefont {Silberberg}},\
  }\href@noop {} {\bibfield  {journal} {\bibinfo  {journal} {{Science}}\
  }\textbf {\bibinfo {volume} {328}},\ \bibinfo {pages} {879} (\bibinfo {year}
  {2010})}\BibitemShut {NoStop}%
\bibitem [{\citenamefont {Giovannetti}\ \emph {et~al.}(2006)\citenamefont
  {Giovannetti}, \citenamefont {Lloyd},\ and\ \citenamefont
  {Maccone}}]{giovannetti04a}%
  \BibitemOpen
  \bibfield  {author} {\bibinfo {author} {\bibfnamefont {V.}~\bibnamefont
  {Giovannetti}}, \bibinfo {author} {\bibfnamefont {S.}~\bibnamefont {Lloyd}},
  \ and\ \bibinfo {author} {\bibfnamefont {L.}~\bibnamefont {Maccone}},\ }\href
  {\doibase 10.1103/PhysRevLett.96.010401} {\bibfield  {journal} {\bibinfo
  {journal} {Phys. Rev. Lett.}\ }\textbf {\bibinfo {volume} {96}},\ \bibinfo
  {pages} {010401} (\bibinfo {year} {2006})}\BibitemShut {NoStop}%
\bibitem [{\citenamefont {D'Angelo}\ \emph {et~al.}(2001)\citenamefont
  {D'Angelo}, \citenamefont {Chekhova},\ and\ \citenamefont
  {Shih}}]{dangelo01a}%
  \BibitemOpen
  \bibfield  {author} {\bibinfo {author} {\bibfnamefont {M.}~\bibnamefont
  {D'Angelo}}, \bibinfo {author} {\bibfnamefont {M.~V.}\ \bibnamefont
  {Chekhova}}, \ and\ \bibinfo {author} {\bibfnamefont {Y.}~\bibnamefont
  {Shih}},\ }\href@noop {} {\bibfield  {journal} {\bibinfo  {journal} {Physical
  review letters}\ }\textbf {\bibinfo {volume} {87}},\ \bibinfo {pages}
  {013602} (\bibinfo {year} {2001})}\BibitemShut {NoStop}%
\bibitem [{\citenamefont {Denk}\ \emph {et~al.}(1990)\citenamefont {Denk},
  \citenamefont {Strickler},\ and\ \citenamefont {Webb}}]{denk90a}%
  \BibitemOpen
  \bibfield  {author} {\bibinfo {author} {\bibfnamefont {W.}~\bibnamefont
  {Denk}}, \bibinfo {author} {\bibfnamefont {J.}~\bibnamefont {Strickler}}, \
  and\ \bibinfo {author} {\bibfnamefont {W.}~\bibnamefont {Webb}},\ }\href@noop
  {} {\bibfield  {journal} {\bibinfo  {journal} {{Science}}\ }\textbf {\bibinfo
  {volume} {248}},\ \bibinfo {pages} {73} (\bibinfo {year} {1990})}\BibitemShut
  {NoStop}%
\bibitem [{\citenamefont {Horton}\ \emph {et~al.}(2013)\citenamefont {Horton},
  \citenamefont {Wang}, \citenamefont {Kobat}, \citenamefont {Clark},
  \citenamefont {Wise},\ and\ \citenamefont {C.~Schaffer}}]{horton13a}%
  \BibitemOpen
  \bibfield  {author} {\bibinfo {author} {\bibfnamefont {N.}~\bibnamefont
  {Horton}}, \bibinfo {author} {\bibfnamefont {D.}~\bibnamefont {Wang}},
  \bibinfo {author} {\bibfnamefont {C.}~\bibnamefont {Kobat}}, \bibinfo
  {author} {\bibfnamefont {F.}~\bibnamefont {Clark}}, \bibinfo {author}
  {\bibfnamefont {C.}~\bibnamefont {Wise}}, \ and\ \bibinfo {author}
  {\bibfnamefont {C.~X.}\ \bibnamefont {C.~Schaffer}},\ }\href@noop {}
  {\bibfield  {journal} {\bibinfo  {journal} {{Nat. Photon.}}\ }\textbf
  {\bibinfo {volume} {7}},\ \bibinfo {pages} {205} (\bibinfo {year}
  {2013})}\BibitemShut {NoStop}%
\bibitem [{\citenamefont {Bruno}\ \emph {et~al.}(2014)\citenamefont {Bruno},
  \citenamefont {Martin}, \citenamefont {Guerreiro}, \citenamefont
  {Sanguinetti},\ and\ \citenamefont {Thew}}]{bruno14a}%
  \BibitemOpen
  \bibfield  {author} {\bibinfo {author} {\bibfnamefont {N.}~\bibnamefont
  {Bruno}}, \bibinfo {author} {\bibfnamefont {A.}~\bibnamefont {Martin}},
  \bibinfo {author} {\bibfnamefont {T.}~\bibnamefont {Guerreiro}}, \bibinfo
  {author} {\bibfnamefont {B.}~\bibnamefont {Sanguinetti}}, \ and\ \bibinfo
  {author} {\bibfnamefont {R.~T.}\ \bibnamefont {Thew}},\ }\href@noop {}
  {\bibfield  {journal} {\bibinfo  {journal} {Optics express}\ }\textbf
  {\bibinfo {volume} {22}},\ \bibinfo {pages} {17246} (\bibinfo {year}
  {2014})}\BibitemShut {NoStop}%
\bibitem [{\citenamefont {Ourjoumtsev}\ \emph {et~al.}(2006)\citenamefont
  {Ourjoumtsev}, \citenamefont {Tualle-Brouri},\ and\ \citenamefont
  {Grangier}}]{ourjoumtsev06b}%
  \BibitemOpen
  \bibfield  {author} {\bibinfo {author} {\bibfnamefont {A.}~\bibnamefont
  {Ourjoumtsev}}, \bibinfo {author} {\bibfnamefont {R.}~\bibnamefont
  {Tualle-Brouri}}, \ and\ \bibinfo {author} {\bibfnamefont {P.}~\bibnamefont
  {Grangier}},\ }\href {\doibase 10.1103/PhysRevLett.96.213601} {\bibfield
  {journal} {\bibinfo  {journal} {Phys. Rev. Lett.}\ }\textbf {\bibinfo
  {volume} {96}},\ \bibinfo {pages} {213601} (\bibinfo {year}
  {2006})}\BibitemShut {NoStop}%
\bibitem [{\citenamefont {Zavatta}\ \emph {et~al.}(2008)\citenamefont
  {Zavatta}, \citenamefont {Parigi},\ and\ \citenamefont
  {Bellini}}]{zavatta08a}%
  \BibitemOpen
  \bibfield  {author} {\bibinfo {author} {\bibfnamefont {A.}~\bibnamefont
  {Zavatta}}, \bibinfo {author} {\bibfnamefont {V.}~\bibnamefont {Parigi}}, \
  and\ \bibinfo {author} {\bibfnamefont {M.}~\bibnamefont {Bellini}},\ }\href
  {\doibase 10.1103/PhysRevA.78.033809} {\bibfield  {journal} {\bibinfo
  {journal} {Phys. Rev. A}\ }\textbf {\bibinfo {volume} {78}},\ \bibinfo
  {pages} {033809} (\bibinfo {year} {2008})}\BibitemShut {NoStop}%
\bibitem [{\citenamefont {Yao}\ \emph {et~al.}(2012)\citenamefont {Yao},
  \citenamefont {Wang}, \citenamefont {Xu}, \citenamefont {Lu}, \citenamefont
  {Pan}, \citenamefont {Bao}, \citenamefont {Peng}, \citenamefont {Lu},
  \citenamefont {Chen},\ and\ \citenamefont {Pan}}]{yao12a}%
  \BibitemOpen
  \bibfield  {author} {\bibinfo {author} {\bibfnamefont {X.-C.}\ \bibnamefont
  {Yao}}, \bibinfo {author} {\bibfnamefont {T.-X.}\ \bibnamefont {Wang}},
  \bibinfo {author} {\bibfnamefont {P.}~\bibnamefont {Xu}}, \bibinfo {author}
  {\bibfnamefont {H.}~\bibnamefont {Lu}}, \bibinfo {author} {\bibfnamefont
  {G.-S.}\ \bibnamefont {Pan}}, \bibinfo {author} {\bibfnamefont {X.-H.}\
  \bibnamefont {Bao}}, \bibinfo {author} {\bibfnamefont {C.-Z.}\ \bibnamefont
  {Peng}}, \bibinfo {author} {\bibfnamefont {C.-Y.}\ \bibnamefont {Lu}},
  \bibinfo {author} {\bibfnamefont {Y.-A.}\ \bibnamefont {Chen}}, \ and\
  \bibinfo {author} {\bibfnamefont {J.-W.}\ \bibnamefont {Pan}},\ }\href@noop
  {} {\bibfield  {journal} {\bibinfo  {journal} {Nature Photonics}\ }\textbf
  {\bibinfo {volume} {6}},\ \bibinfo {pages} {225} (\bibinfo {year}
  {2012})}\BibitemShut {NoStop}%
\bibitem [{\citenamefont {Law}\ and\ \citenamefont {Eberly}(1996)}]{law96a}%
  \BibitemOpen
  \bibfield  {author} {\bibinfo {author} {\bibfnamefont {C.~K.}\ \bibnamefont
  {Law}}\ and\ \bibinfo {author} {\bibfnamefont {J.~H.}\ \bibnamefont
  {Eberly}},\ }\href@noop {} {\bibfield  {journal} {\bibinfo  {journal} {Phys.
  Rev. Lett.}\ }\textbf {\bibinfo {volume} {76}},\ \bibinfo {pages} {1055}
  (\bibinfo {year} {1996})}\BibitemShut {NoStop}%
\bibitem [{\citenamefont {Callsen}\ \emph {et~al.}(2013)\citenamefont
  {Callsen}, \citenamefont {Carmele}, \citenamefont {H\"onig}, \citenamefont
  {Kindel}, \citenamefont {Brunnmeier}, \citenamefont {Wagner}, \citenamefont
  {Stock}, \citenamefont {Reparaz}, \citenamefont {Schliwa}, \citenamefont
  {Reitzenstein}, \citenamefont {Knorr}, \citenamefont {Hoffmann},
  \citenamefont {Kako},\ and\ \citenamefont {Arakawa}}]{callsen13a}%
  \BibitemOpen
  \bibfield  {author} {\bibinfo {author} {\bibfnamefont {G.}~\bibnamefont
  {Callsen}}, \bibinfo {author} {\bibfnamefont {A.}~\bibnamefont {Carmele}},
  \bibinfo {author} {\bibfnamefont {G.}~\bibnamefont {H\"onig}}, \bibinfo
  {author} {\bibfnamefont {C.}~\bibnamefont {Kindel}}, \bibinfo {author}
  {\bibfnamefont {J.}~\bibnamefont {Brunnmeier}}, \bibinfo {author}
  {\bibfnamefont {M.~R.}\ \bibnamefont {Wagner}}, \bibinfo {author}
  {\bibfnamefont {E.}~\bibnamefont {Stock}}, \bibinfo {author} {\bibfnamefont
  {J.~S.}\ \bibnamefont {Reparaz}}, \bibinfo {author} {\bibfnamefont
  {A.}~\bibnamefont {Schliwa}}, \bibinfo {author} {\bibfnamefont
  {S.}~\bibnamefont {Reitzenstein}}, \bibinfo {author} {\bibfnamefont
  {A.}~\bibnamefont {Knorr}}, \bibinfo {author} {\bibfnamefont
  {A.}~\bibnamefont {Hoffmann}}, \bibinfo {author} {\bibfnamefont
  {S.}~\bibnamefont {Kako}}, \ and\ \bibinfo {author} {\bibfnamefont
  {Y.}~\bibnamefont {Arakawa}},\ }\href {\doibase 10.1103/PhysRevB.87.245314}
  {\bibfield  {journal} {\bibinfo  {journal} {Phys. Rev. B}\ }\textbf {\bibinfo
  {volume} {87}},\ \bibinfo {pages} {245314} (\bibinfo {year}
  {2013})}\BibitemShut {NoStop}%
\bibitem [{\citenamefont {Dousse}\ \emph {et~al.}(2010)\citenamefont {Dousse},
  \citenamefont {Suffczy{\'n}ski}, \citenamefont {Beveratos}, \citenamefont
  {Krebs}, \citenamefont {Lema{\^\i}tre}, \citenamefont {Sagnes}, \citenamefont
  {Bloch}, \citenamefont {Voisin},\ and\ \citenamefont
  {Senellart}}]{dousse10a}%
  \BibitemOpen
  \bibfield  {author} {\bibinfo {author} {\bibfnamefont {A.}~\bibnamefont
  {Dousse}}, \bibinfo {author} {\bibfnamefont {J.}~\bibnamefont
  {Suffczy{\'n}ski}}, \bibinfo {author} {\bibfnamefont {A.}~\bibnamefont
  {Beveratos}}, \bibinfo {author} {\bibfnamefont {O.}~\bibnamefont {Krebs}},
  \bibinfo {author} {\bibfnamefont {A.}~\bibnamefont {Lema{\^\i}tre}}, \bibinfo
  {author} {\bibfnamefont {I.}~\bibnamefont {Sagnes}}, \bibinfo {author}
  {\bibfnamefont {J.}~\bibnamefont {Bloch}}, \bibinfo {author} {\bibfnamefont
  {P.}~\bibnamefont {Voisin}}, \ and\ \bibinfo {author} {\bibfnamefont
  {P.}~\bibnamefont {Senellart}},\ }\href@noop {} {\bibfield  {journal}
  {\bibinfo  {journal} {{Nature}}\ }\textbf {\bibinfo {volume} {466}},\
  \bibinfo {pages} {217} (\bibinfo {year} {2010})}\BibitemShut {NoStop}%
\bibitem [{\citenamefont {M{\"u}ller}\ \emph {et~al.}(2014)\citenamefont
  {M{\"u}ller}, \citenamefont {Bounouar}, \citenamefont {J{\"o}ns},
  \citenamefont {Gl{\"a}ssl},\ and\ \citenamefont {Michler}}]{muller14a}%
  \BibitemOpen
  \bibfield  {author} {\bibinfo {author} {\bibfnamefont {M.}~\bibnamefont
  {M{\"u}ller}}, \bibinfo {author} {\bibfnamefont {S.}~\bibnamefont
  {Bounouar}}, \bibinfo {author} {\bibfnamefont {K.~D.}\ \bibnamefont
  {J{\"o}ns}}, \bibinfo {author} {\bibfnamefont {M.}~\bibnamefont
  {Gl{\"a}ssl}}, \ and\ \bibinfo {author} {\bibfnamefont {P.}~\bibnamefont
  {Michler}},\ }\href@noop {} {\bibfield  {journal} {\bibinfo  {journal}
  {Nature Photonics}\ }\textbf {\bibinfo {volume} {8}},\ \bibinfo {pages} {224}
  (\bibinfo {year} {2014})}\BibitemShut {NoStop}%
\bibitem [{\citenamefont {Hofheinz}\ \emph {et~al.}(2008)\citenamefont
  {Hofheinz}, \citenamefont {Weig}, \citenamefont {Ansmann}, \citenamefont
  {Bialczak}, \citenamefont {Lucero}, \citenamefont {Neeley}, \citenamefont
  {O'Connell}, \citenamefont {Wang}, \citenamefont {Martinis1},\ and\
  \citenamefont {Cleland}}]{hofheinz08a}%
  \BibitemOpen
  \bibfield  {author} {\bibinfo {author} {\bibfnamefont {M.}~\bibnamefont
  {Hofheinz}}, \bibinfo {author} {\bibfnamefont {E.}~\bibnamefont {Weig}},
  \bibinfo {author} {\bibfnamefont {M.}~\bibnamefont {Ansmann}}, \bibinfo
  {author} {\bibfnamefont {R.}~\bibnamefont {Bialczak}}, \bibinfo {author}
  {\bibfnamefont {E.}~\bibnamefont {Lucero}}, \bibinfo {author} {\bibfnamefont
  {M.}~\bibnamefont {Neeley}}, \bibinfo {author} {\bibfnamefont
  {A.}~\bibnamefont {O'Connell}}, \bibinfo {author} {\bibfnamefont
  {H.}~\bibnamefont {Wang}}, \bibinfo {author} {\bibfnamefont {J.}~\bibnamefont
  {Martinis1}}, \ and\ \bibinfo {author} {\bibfnamefont {A.}~\bibnamefont
  {Cleland}},\ }\href@noop {} {\bibfield  {journal} {\bibinfo  {journal}
  {{Nature}}\ }\textbf {\bibinfo {volume} {454}},\ \bibinfo {pages} {310}
  (\bibinfo {year} {2008})}\BibitemShut {NoStop}%
\bibitem [{\citenamefont {Hofheinz}\ \emph {et~al.}(2009)\citenamefont
  {Hofheinz}, \citenamefont {Wang}, \citenamefont {Ansmann}, \citenamefont
  {Bialczak}, \citenamefont {Lucero}, \citenamefont {Neeley}, \citenamefont
  {O'Connell}, \citenamefont {Sank}, \citenamefont {Wenner}, \citenamefont
  {Martinis} \emph {et~al.}}]{hofheinz09a}%
  \BibitemOpen
  \bibfield  {author} {\bibinfo {author} {\bibfnamefont {M.}~\bibnamefont
  {Hofheinz}}, \bibinfo {author} {\bibfnamefont {H.}~\bibnamefont {Wang}},
  \bibinfo {author} {\bibfnamefont {M.}~\bibnamefont {Ansmann}}, \bibinfo
  {author} {\bibfnamefont {R.~C.}\ \bibnamefont {Bialczak}}, \bibinfo {author}
  {\bibfnamefont {E.}~\bibnamefont {Lucero}}, \bibinfo {author} {\bibfnamefont
  {M.}~\bibnamefont {Neeley}}, \bibinfo {author} {\bibfnamefont
  {A.}~\bibnamefont {O'Connell}}, \bibinfo {author} {\bibfnamefont
  {D.}~\bibnamefont {Sank}}, \bibinfo {author} {\bibfnamefont {J.}~\bibnamefont
  {Wenner}}, \bibinfo {author} {\bibfnamefont {J.~M.}\ \bibnamefont
  {Martinis}},  \emph {et~al.},\ }\href@noop {} {\bibfield  {journal} {\bibinfo
   {journal} {Nature}\ }\textbf {\bibinfo {volume} {459}},\ \bibinfo {pages}
  {546} (\bibinfo {year} {2009})}\BibitemShut {NoStop}%
\bibitem [{\citenamefont {Johnson}\ \emph {et~al.}(2010)\citenamefont
  {Johnson}, \citenamefont {Reed}, \citenamefont {Houck}, \citenamefont
  {Schuster}, \citenamefont {Bishop}, \citenamefont {Ginossar}, \citenamefont
  {Gambetta}, \citenamefont {DiCarlo}, \citenamefont {Frunzio}, \citenamefont
  {Girvin} \emph {et~al.}}]{johnson10a}%
  \BibitemOpen
  \bibfield  {author} {\bibinfo {author} {\bibfnamefont {B.}~\bibnamefont
  {Johnson}}, \bibinfo {author} {\bibfnamefont {M.}~\bibnamefont {Reed}},
  \bibinfo {author} {\bibfnamefont {A.}~\bibnamefont {Houck}}, \bibinfo
  {author} {\bibfnamefont {D.}~\bibnamefont {Schuster}}, \bibinfo {author}
  {\bibfnamefont {L.~S.}\ \bibnamefont {Bishop}}, \bibinfo {author}
  {\bibfnamefont {E.}~\bibnamefont {Ginossar}}, \bibinfo {author}
  {\bibfnamefont {J.}~\bibnamefont {Gambetta}}, \bibinfo {author}
  {\bibfnamefont {L.}~\bibnamefont {DiCarlo}}, \bibinfo {author} {\bibfnamefont
  {L.}~\bibnamefont {Frunzio}}, \bibinfo {author} {\bibfnamefont
  {S.}~\bibnamefont {Girvin}},  \emph {et~al.},\ }\href@noop {} {\bibfield
  {journal} {\bibinfo  {journal} {Nature Physics}\ }\textbf {\bibinfo {volume}
  {6}},\ \bibinfo {pages} {663} (\bibinfo {year} {2010})}\BibitemShut {NoStop}%
\bibitem [{\citenamefont {Bozyigit}\ \emph {et~al.}(2011)\citenamefont
  {Bozyigit}, \citenamefont {Lang}, \citenamefont {Steffen}, \citenamefont
  {Fink}, \citenamefont {Eichler}, \citenamefont {Baur}, \citenamefont
  {Bianchetti}, \citenamefont {Leek}, \citenamefont {Filipp}, \citenamefont
  {da~Silva}, \citenamefont {Blais},\ and\ \citenamefont
  {Wallraff}}]{bozyigit11a}%
  \BibitemOpen
  \bibfield  {author} {\bibinfo {author} {\bibfnamefont {D.}~\bibnamefont
  {Bozyigit}}, \bibinfo {author} {\bibfnamefont {C.}~\bibnamefont {Lang}},
  \bibinfo {author} {\bibfnamefont {L.}~\bibnamefont {Steffen}}, \bibinfo
  {author} {\bibfnamefont {J.~M.}\ \bibnamefont {Fink}}, \bibinfo {author}
  {\bibfnamefont {C.}~\bibnamefont {Eichler}}, \bibinfo {author} {\bibfnamefont
  {M.}~\bibnamefont {Baur}}, \bibinfo {author} {\bibfnamefont {R.}~\bibnamefont
  {Bianchetti}}, \bibinfo {author} {\bibfnamefont {P.~J.}\ \bibnamefont
  {Leek}}, \bibinfo {author} {\bibfnamefont {S.}~\bibnamefont {Filipp}},
  \bibinfo {author} {\bibfnamefont {M.~P.}\ \bibnamefont {da~Silva}}, \bibinfo
  {author} {\bibfnamefont {A.}~\bibnamefont {Blais}}, \ and\ \bibinfo {author}
  {\bibfnamefont {A.}~\bibnamefont {Wallraff}},\ }\href@noop {} {\bibfield
  {journal} {\bibinfo  {journal} {{Nat. Phys.}}\ }\textbf {\bibinfo {volume}
  {7}},\ \bibinfo {pages} {154} (\bibinfo {year} {2011})}\BibitemShut {NoStop}%
\bibitem [{\citenamefont {Eichler}\ \emph {et~al.}(2011)\citenamefont
  {Eichler}, \citenamefont {Bozyigit}, \citenamefont {Lang}, \citenamefont
  {Steffen}, \citenamefont {Fink},\ and\ \citenamefont
  {Wallraff}}]{eichler11a}%
  \BibitemOpen
  \bibfield  {author} {\bibinfo {author} {\bibfnamefont {C.}~\bibnamefont
  {Eichler}}, \bibinfo {author} {\bibfnamefont {D.}~\bibnamefont {Bozyigit}},
  \bibinfo {author} {\bibfnamefont {C.}~\bibnamefont {Lang}}, \bibinfo {author}
  {\bibfnamefont {L.}~\bibnamefont {Steffen}}, \bibinfo {author} {\bibfnamefont
  {J.}~\bibnamefont {Fink}}, \ and\ \bibinfo {author} {\bibfnamefont
  {A.}~\bibnamefont {Wallraff}},\ }\href {\doibase
  10.1103/PhysRevLett.106.220503} {\bibfield  {journal} {\bibinfo  {journal}
  {Phys. Rev. Lett.}\ }\textbf {\bibinfo {volume} {106}},\ \bibinfo {pages}
  {220503} (\bibinfo {year} {2011})}\BibitemShut {NoStop}%
\bibitem [{\citenamefont {Pechal}\ \emph {et~al.}(2014)\citenamefont {Pechal},
  \citenamefont {Huthmacher}, \citenamefont {Eichler}, \citenamefont
  {Zeytino\ifmmode~\breve{g}\else \u{g}\fi{}lu}, \citenamefont {Abdumalikov},
  \citenamefont {Berger}, \citenamefont {Wallraff},\ and\ \citenamefont
  {Filipp}}]{pechal14a}%
  \BibitemOpen
  \bibfield  {author} {\bibinfo {author} {\bibfnamefont {M.}~\bibnamefont
  {Pechal}}, \bibinfo {author} {\bibfnamefont {L.}~\bibnamefont {Huthmacher}},
  \bibinfo {author} {\bibfnamefont {C.}~\bibnamefont {Eichler}}, \bibinfo
  {author} {\bibfnamefont {S.}~\bibnamefont {Zeytino\ifmmode~\breve{g}\else
  \u{g}\fi{}lu}}, \bibinfo {author} {\bibfnamefont {A.~A.}\ \bibnamefont
  {Abdumalikov}}, \bibinfo {author} {\bibfnamefont {S.}~\bibnamefont {Berger}},
  \bibinfo {author} {\bibfnamefont {A.}~\bibnamefont {Wallraff}}, \ and\
  \bibinfo {author} {\bibfnamefont {S.}~\bibnamefont {Filipp}},\ }\href
  {\doibase 10.1103/PhysRevX.4.041010} {\bibfield  {journal} {\bibinfo
  {journal} {Phys. Rev. X}\ }\textbf {\bibinfo {volume} {4}},\ \bibinfo {pages}
  {041010} (\bibinfo {year} {2014})}\BibitemShut {NoStop}%
\bibitem [{\citenamefont {{Sanchez Mu\~noz}}\ \emph {et~al.}(2014)\citenamefont
  {{Sanchez Mu\~noz}}, \citenamefont {del Valle}, \citenamefont {Tudela},
  \citenamefont {M\"uller}, \citenamefont {Lichtmannecker}, \citenamefont
  {Kaniber}, \citenamefont {Tejedor}, \citenamefont {Finley},\ and\
  \citenamefont {Laussy}}]{sanchezmunoz14a}%
  \BibitemOpen
  \bibfield  {author} {\bibinfo {author} {\bibfnamefont {C.}~\bibnamefont
  {{Sanchez Mu\~noz}}}, \bibinfo {author} {\bibfnamefont {E.}~\bibnamefont {del
  Valle}}, \bibinfo {author} {\bibfnamefont {A.~G.}\ \bibnamefont {Tudela}},
  \bibinfo {author} {\bibfnamefont {K.}~\bibnamefont {M\"uller}}, \bibinfo
  {author} {\bibfnamefont {S.}~\bibnamefont {Lichtmannecker}}, \bibinfo
  {author} {\bibfnamefont {M.}~\bibnamefont {Kaniber}}, \bibinfo {author}
  {\bibfnamefont {C.}~\bibnamefont {Tejedor}}, \bibinfo {author} {\bibfnamefont
  {J.}~\bibnamefont {Finley}}, \ and\ \bibinfo {author} {\bibfnamefont
  {F.}~\bibnamefont {Laussy}},\ }\href@noop {} {\bibfield  {journal} {\bibinfo
  {journal} {{Nat. Photon.}}\ }\textbf {\bibinfo {volume} {8}},\ \bibinfo
  {pages} {550} (\bibinfo {year} {2014})}\BibitemShut {NoStop}%
\bibitem [{\citenamefont {S{\'a}nchez-Mu{\~n}oz}\ \emph
  {et~al.}(2015)\citenamefont {S{\'a}nchez-Mu{\~n}oz}, \citenamefont {Laussy},
  \citenamefont {Tejedor},\ and\ \citenamefont {del Valle}}]{sanchezmunoz15a}%
  \BibitemOpen
  \bibfield  {author} {\bibinfo {author} {\bibfnamefont {C.}~\bibnamefont
  {S{\'a}nchez-Mu{\~n}oz}}, \bibinfo {author} {\bibfnamefont {F.~P.}\
  \bibnamefont {Laussy}}, \bibinfo {author} {\bibfnamefont {C.}~\bibnamefont
  {Tejedor}}, \ and\ \bibinfo {author} {\bibfnamefont {E.}~\bibnamefont {del
  Valle}},\ }\href@noop {} {\bibfield  {journal} {\bibinfo  {journal}
  {arXiv:1506.05050}\ } (\bibinfo {year} {2015})}\BibitemShut {NoStop}%
\bibitem [{\citenamefont {Gonz\'alez-Tudela}\ \emph {et~al.}(2015)\citenamefont
  {Gonz\'alez-Tudela}, \citenamefont {Paulisch}, \citenamefont {Chang},
  \citenamefont {Kimble},\ and\ \citenamefont {Cirac}}]{gonzaleztudela15d}%
  \BibitemOpen
  \bibfield  {author} {\bibinfo {author} {\bibfnamefont {A.}~\bibnamefont
  {Gonz\'alez-Tudela}}, \bibinfo {author} {\bibfnamefont {V.}~\bibnamefont
  {Paulisch}}, \bibinfo {author} {\bibfnamefont {D.~E.}\ \bibnamefont {Chang}},
  \bibinfo {author} {\bibfnamefont {H.~J.}\ \bibnamefont {Kimble}}, \ and\
  \bibinfo {author} {\bibfnamefont {J.~I.}\ \bibnamefont {Cirac}},\ }\href
  {\doibase 10.1103/PhysRevLett.115.163603} {\bibfield  {journal} {\bibinfo
  {journal} {Phys. Rev. Lett.}\ }\textbf {\bibinfo {volume} {115}},\ \bibinfo
  {pages} {163603} (\bibinfo {year} {2015})}\BibitemShut {NoStop}%
\bibitem [{\citenamefont {Rundquist}\ \emph {et~al.}(2014)\citenamefont
  {Rundquist}, \citenamefont {Bajcsy}, \citenamefont {Majumdar}, \citenamefont
  {Sarmiento}, \citenamefont {Fischer}, \citenamefont {Lagoudakis},
  \citenamefont {Buckley}, \citenamefont {Piggott},\ and\ \citenamefont
  {Vu\ifmmode \check{c}\else \v{c}\fi{}kovi\ifmmode~\acute{c}\else
  \'{c}\fi{}}}]{rundquist14a}%
  \BibitemOpen
  \bibfield  {author} {\bibinfo {author} {\bibfnamefont {A.}~\bibnamefont
  {Rundquist}}, \bibinfo {author} {\bibfnamefont {M.}~\bibnamefont {Bajcsy}},
  \bibinfo {author} {\bibfnamefont {A.}~\bibnamefont {Majumdar}}, \bibinfo
  {author} {\bibfnamefont {T.}~\bibnamefont {Sarmiento}}, \bibinfo {author}
  {\bibfnamefont {K.}~\bibnamefont {Fischer}}, \bibinfo {author} {\bibfnamefont
  {K.~G.}\ \bibnamefont {Lagoudakis}}, \bibinfo {author} {\bibfnamefont
  {S.}~\bibnamefont {Buckley}}, \bibinfo {author} {\bibfnamefont {A.~Y.}\
  \bibnamefont {Piggott}}, \ and\ \bibinfo {author} {\bibfnamefont
  {J.}~\bibnamefont {Vu\ifmmode \check{c}\else
  \v{c}\fi{}kovi\ifmmode~\acute{c}\else \'{c}\fi{}}},\ }\href {\doibase
  10.1103/PhysRevA.90.023846} {\bibfield  {journal} {\bibinfo  {journal} {Phys.
  Rev. A}\ }\textbf {\bibinfo {volume} {90}},\ \bibinfo {pages} {023846}
  (\bibinfo {year} {2014})}\BibitemShut {NoStop}%
\bibitem [{\citenamefont {Koch}\ \emph {et~al.}(2011)\citenamefont {Koch},
  \citenamefont {Sames}, \citenamefont {Balbach}, \citenamefont {Chibani},
  \citenamefont {Kubanek}, \citenamefont {Murr}, \citenamefont {Wilk},\ and\
  \citenamefont {Rempe}}]{koch11a}%
  \BibitemOpen
  \bibfield  {author} {\bibinfo {author} {\bibfnamefont {M.}~\bibnamefont
  {Koch}}, \bibinfo {author} {\bibfnamefont {C.}~\bibnamefont {Sames}},
  \bibinfo {author} {\bibfnamefont {M.}~\bibnamefont {Balbach}}, \bibinfo
  {author} {\bibfnamefont {H.}~\bibnamefont {Chibani}}, \bibinfo {author}
  {\bibfnamefont {A.}~\bibnamefont {Kubanek}}, \bibinfo {author} {\bibfnamefont
  {K.}~\bibnamefont {Murr}}, \bibinfo {author} {\bibfnamefont {T.}~\bibnamefont
  {Wilk}}, \ and\ \bibinfo {author} {\bibfnamefont {G.}~\bibnamefont {Rempe}},\
  }\href@noop {} {\bibfield  {journal} {\bibinfo  {journal} {{Phys. Rev.
  Lett.}}\ }\textbf {\bibinfo {volume} {107}},\ \bibinfo {pages} {023601}
  (\bibinfo {year} {2011})}\BibitemShut {NoStop}%
\bibitem [{\citenamefont {Hong}\ \emph {et~al.}(2010)\citenamefont {Hong},
  \citenamefont {Nha}, \citenamefont {Lee},\ and\ \citenamefont
  {An}}]{hong10a}%
  \BibitemOpen
  \bibfield  {author} {\bibinfo {author} {\bibfnamefont {H.-G.}\ \bibnamefont
  {Hong}}, \bibinfo {author} {\bibfnamefont {H.}~\bibnamefont {Nha}}, \bibinfo
  {author} {\bibfnamefont {J.-H.}\ \bibnamefont {Lee}}, \ and\ \bibinfo
  {author} {\bibfnamefont {K.}~\bibnamefont {An}},\ }\href@noop {} {\bibfield
  {journal} {\bibinfo  {journal} {{Opt. Express}}\ }\textbf {\bibinfo {volume}
  {18}},\ \bibinfo {pages} {7092} (\bibinfo {year} {2010})}\BibitemShut
  {NoStop}%
\bibitem [{\citenamefont {Le~Boit\'e}\ \emph {et~al.}(2014)\citenamefont
  {Le~Boit\'e}, \citenamefont {Orso},\ and\ \citenamefont
  {Ciuti}}]{leboite14a}%
  \BibitemOpen
  \bibfield  {author} {\bibinfo {author} {\bibfnamefont {A.}~\bibnamefont
  {Le~Boit\'e}}, \bibinfo {author} {\bibfnamefont {G.}~\bibnamefont {Orso}}, \
  and\ \bibinfo {author} {\bibfnamefont {C.}~\bibnamefont {Ciuti}},\ }\href
  {\doibase 10.1103/PhysRevA.90.063821} {\bibfield  {journal} {\bibinfo
  {journal} {Phys. Rev. A}\ }\textbf {\bibinfo {volume} {90}},\ \bibinfo
  {pages} {063821} (\bibinfo {year} {2014})}\BibitemShut {NoStop}%
\bibitem [{\citenamefont {Shi}\ and\ \citenamefont {Sun}(2009)}]{shi09a}%
  \BibitemOpen
  \bibfield  {author} {\bibinfo {author} {\bibfnamefont {T.}~\bibnamefont
  {Shi}}\ and\ \bibinfo {author} {\bibfnamefont {C.}~\bibnamefont {Sun}},\
  }\href@noop {} {\bibfield  {journal} {\bibinfo  {journal} {Physical Review
  B}\ }\textbf {\bibinfo {volume} {79}},\ \bibinfo {pages} {205111} (\bibinfo
  {year} {2009})}\BibitemShut {NoStop}%
\bibitem [{\citenamefont {Caneva}\ \emph {et~al.}()\citenamefont {Caneva},
  \citenamefont {Manzoni}, \citenamefont {Shi}, \citenamefont {Douglas},
  \citenamefont {Cirac},\ and\ \citenamefont {Chang}}]{caneva15a}%
  \BibitemOpen
  \bibfield  {author} {\bibinfo {author} {\bibfnamefont {T.}~\bibnamefont
  {Caneva}}, \bibinfo {author} {\bibfnamefont {M.~T.}\ \bibnamefont {Manzoni}},
  \bibinfo {author} {\bibfnamefont {T.}~\bibnamefont {Shi}}, \bibinfo {author}
  {\bibfnamefont {J.~S.}\ \bibnamefont {Douglas}}, \bibinfo {author}
  {\bibfnamefont {J.~I.}\ \bibnamefont {Cirac}}, \ and\ \bibinfo {author}
  {\bibfnamefont {D.~E.}\ \bibnamefont {Chang}},\ }\href
  {http://arxiv.org/abs/1501.04427} {\ }\Eprint {http://arxiv.org/abs/arXiv:
  1501.04427} {arXiv: 1501.04427} \BibitemShut {NoStop}%
\bibitem [{\citenamefont {Shi}\ \emph {et~al.}(2015)\citenamefont {Shi},
  \citenamefont {Chang},\ and\ \citenamefont {Cirac}}]{shi15a}%
  \BibitemOpen
  \bibfield  {author} {\bibinfo {author} {\bibfnamefont {T.}~\bibnamefont
  {Shi}}, \bibinfo {author} {\bibfnamefont {D.~E.}\ \bibnamefont {Chang}}, \
  and\ \bibinfo {author} {\bibfnamefont {J.~I.}\ \bibnamefont {Cirac}},\
  }\href@noop {} {\bibfield  {journal} {\bibinfo  {journal} {arXiv:1507.08699}\
  } (\bibinfo {year} {2015})}\BibitemShut {NoStop}%
\bibitem [{\citenamefont {Sakurai}\ and\ \citenamefont
  {Napolitano}(2014)}]{sakurai14a}%
  \BibitemOpen
  \bibfield  {author} {\bibinfo {author} {\bibfnamefont {J.~J.}\ \bibnamefont
  {Sakurai}}\ and\ \bibinfo {author} {\bibfnamefont {J.~J.}\ \bibnamefont
  {Napolitano}},\ }\href@noop {} {\emph {\bibinfo {title} {Modern quantum
  mechanics}}}\ (\bibinfo  {publisher} {Pearson Higher Ed},\ \bibinfo {year}
  {2014})\BibitemShut {NoStop}%
\bibitem [{sup()}]{supmat}%
  \BibitemOpen
  \href@noop {} {}\bibinfo {note} {See the supplemental material accompanying
  this letter, where we introduce the scattering formalism (explaining how to
  obtain scattering amplitudes and wavefunctions, and connecting these with
  multi-time correlation functions of weakly driven systems), analyze the
  timescales our system, providing approximate analytical expressions for its
  relevant observables, and prove in detail how the suggested circuit QED
  architecture gives rise to the desired system Hamiltonian.}\BibitemShut
  {Stop}%
\bibitem [{\citenamefont {Gardiner}\ and\ \citenamefont
  {Zoller}(2000)}]{gardiner_book00a}%
  \BibitemOpen
  \bibfield  {author} {\bibinfo {author} {\bibfnamefont {G.~W.}\ \bibnamefont
  {Gardiner}}\ and\ \bibinfo {author} {\bibfnamefont {P.}~\bibnamefont
  {Zoller}},\ }\href@noop {} {\emph {\bibinfo {title} {{Quantum Noise}}}},\
  \bibinfo {edition} {2nd}\ ed.\ (\bibinfo  {publisher} {Springer-Verlag,
  Berlin},\ \bibinfo {year} {2000})\BibitemShut {NoStop}%
\bibitem [{\citenamefont {Carmichael}(2002)}]{carmichael_book02a}%
  \BibitemOpen
  \bibfield  {author} {\bibinfo {author} {\bibfnamefont {H.~J.}\ \bibnamefont
  {Carmichael}},\ }\href@noop {} {\emph {\bibinfo {title} {{Statistical methods
  in quantum optics 1}}}},\ \bibinfo {edition} {2nd}\ ed.\ (\bibinfo
  {publisher} {Springer},\ \bibinfo {year} {2002})\BibitemShut {NoStop}%
\bibitem [{\citenamefont {Liew}\ and\ \citenamefont {Savona}(2010)}]{liew10a}%
  \BibitemOpen
  \bibfield  {author} {\bibinfo {author} {\bibfnamefont {T.~C.~H.}\
  \bibnamefont {Liew}}\ and\ \bibinfo {author} {\bibfnamefont {V.}~\bibnamefont
  {Savona}},\ }\href {\doibase 10.1103/PhysRevLett.104.183601} {\bibfield
  {journal} {\bibinfo  {journal} {Phys. Rev. Lett.}\ }\textbf {\bibinfo
  {volume} {104}},\ \bibinfo {pages} {183601} (\bibinfo {year}
  {2010})}\BibitemShut {NoStop}%
\bibitem [{\citenamefont {Bamba}\ \emph {et~al.}(2011)\citenamefont {Bamba},
  \citenamefont {Imamo\ifmmode~\breve{g}\else \u{g}\fi{}lu}, \citenamefont
  {Carusotto},\ and\ \citenamefont {Ciuti}}]{bamba11a}%
  \BibitemOpen
  \bibfield  {author} {\bibinfo {author} {\bibfnamefont {M.}~\bibnamefont
  {Bamba}}, \bibinfo {author} {\bibfnamefont {A.}~\bibnamefont
  {Imamo\ifmmode~\breve{g}\else \u{g}\fi{}lu}}, \bibinfo {author}
  {\bibfnamefont {I.}~\bibnamefont {Carusotto}}, \ and\ \bibinfo {author}
  {\bibfnamefont {C.}~\bibnamefont {Ciuti}},\ }\href {\doibase
  10.1103/PhysRevA.83.021802} {\bibfield  {journal} {\bibinfo  {journal} {Phys.
  Rev. A}\ }\textbf {\bibinfo {volume} {83}},\ \bibinfo {pages} {021802}
  (\bibinfo {year} {2011})}\BibitemShut {NoStop}%
\bibitem [{\citenamefont {Majumdar}\ \emph {et~al.}(2012)\citenamefont
  {Majumdar}, \citenamefont {Bajcsy}, \citenamefont {Rundquist},\ and\
  \citenamefont {Vu\ifmmode \check{c}\else
  \v{c}\fi{}kovi\ifmmode~\acute{c}\else \'{c}\fi{}}}]{majumdar12a}%
  \BibitemOpen
  \bibfield  {author} {\bibinfo {author} {\bibfnamefont {A.}~\bibnamefont
  {Majumdar}}, \bibinfo {author} {\bibfnamefont {M.}~\bibnamefont {Bajcsy}},
  \bibinfo {author} {\bibfnamefont {A.}~\bibnamefont {Rundquist}}, \ and\
  \bibinfo {author} {\bibfnamefont {J.}~\bibnamefont {Vu\ifmmode \check{c}\else
  \v{c}\fi{}kovi\ifmmode~\acute{c}\else \'{c}\fi{}}},\ }\href {\doibase
  10.1103/PhysRevLett.108.183601} {\bibfield  {journal} {\bibinfo  {journal}
  {Phys. Rev. Lett.}\ }\textbf {\bibinfo {volume} {108}},\ \bibinfo {pages}
  {183601} (\bibinfo {year} {2012})}\BibitemShut {NoStop}%
\bibitem [{\citenamefont {Shi}\ and\ \citenamefont {Fan}(2013)}]{shi13a}%
  \BibitemOpen
  \bibfield  {author} {\bibinfo {author} {\bibfnamefont {T.}~\bibnamefont
  {Shi}}\ and\ \bibinfo {author} {\bibfnamefont {S.}~\bibnamefont {Fan}},\
  }\href {\doibase 10.1103/PhysRevA.87.063818} {\bibfield  {journal} {\bibinfo
  {journal} {Phys. Rev. A}\ }\textbf {\bibinfo {volume} {87}},\ \bibinfo
  {pages} {063818} (\bibinfo {year} {2013})}\BibitemShut {NoStop}%
\bibitem [{\citenamefont {Xiang}\ \emph {et~al.}(2013)\citenamefont {Xiang},
  \citenamefont {Ashhab}, \citenamefont {You},\ and\ \citenamefont
  {Nori}}]{xiang13a}%
  \BibitemOpen
  \bibfield  {author} {\bibinfo {author} {\bibfnamefont {Z.-L.}\ \bibnamefont
  {Xiang}}, \bibinfo {author} {\bibfnamefont {S.}~\bibnamefont {Ashhab}},
  \bibinfo {author} {\bibfnamefont {J.~Q.}\ \bibnamefont {You}}, \ and\
  \bibinfo {author} {\bibfnamefont {F.}~\bibnamefont {Nori}},\ }\href {\doibase
  10.1103/RevModPhys.85.623} {\bibfield  {journal} {\bibinfo  {journal} {Rev.
  Mod. Phys.}\ }\textbf {\bibinfo {volume} {85}},\ \bibinfo {pages} {623}
  (\bibinfo {year} {2013})}\BibitemShut {NoStop}%
\bibitem [{\citenamefont {Devoret}\ and\ \citenamefont
  {Schoelkopf}(2013)}]{devoret13a}%
  \BibitemOpen
  \bibfield  {author} {\bibinfo {author} {\bibfnamefont {M.}~\bibnamefont
  {Devoret}}\ and\ \bibinfo {author} {\bibfnamefont {R.}~\bibnamefont
  {Schoelkopf}},\ }\href@noop {} {\bibfield  {journal} {\bibinfo  {journal}
  {Science}\ }\textbf {\bibinfo {volume} {339}},\ \bibinfo {pages} {1169}
  (\bibinfo {year} {2013})}\BibitemShut {NoStop}%
\bibitem [{\citenamefont {Schoelkopf}\ and\ \citenamefont
  {Girvin}(2008)}]{schoelkopf08a}%
  \BibitemOpen
  \bibfield  {author} {\bibinfo {author} {\bibfnamefont {R.}~\bibnamefont
  {Schoelkopf}}\ and\ \bibinfo {author} {\bibfnamefont {S.}~\bibnamefont
  {Girvin}},\ }\href@noop {} {\bibfield  {journal} {\bibinfo  {journal}
  {Nature}\ }\textbf {\bibinfo {volume} {451}},\ \bibinfo {pages} {664}
  (\bibinfo {year} {2008})}\BibitemShut {NoStop}%
\bibitem [{\citenamefont {Koch}\ \emph {et~al.}(2007)\citenamefont {Koch},
  \citenamefont {Yu}, \citenamefont {Gambetta}, \citenamefont {Houck},
  \citenamefont {Schuster}, \citenamefont {Majer}, \citenamefont {Blais},
  \citenamefont {Devoret}, \citenamefont {Girvin},\ and\ \citenamefont
  {Schoelkopf}}]{koch07a}%
  \BibitemOpen
  \bibfield  {author} {\bibinfo {author} {\bibfnamefont {J.}~\bibnamefont
  {Koch}}, \bibinfo {author} {\bibfnamefont {T.~M.}\ \bibnamefont {Yu}},
  \bibinfo {author} {\bibfnamefont {J.}~\bibnamefont {Gambetta}}, \bibinfo
  {author} {\bibfnamefont {A.~A.}\ \bibnamefont {Houck}}, \bibinfo {author}
  {\bibfnamefont {D.~I.}\ \bibnamefont {Schuster}}, \bibinfo {author}
  {\bibfnamefont {J.}~\bibnamefont {Majer}}, \bibinfo {author} {\bibfnamefont
  {A.}~\bibnamefont {Blais}}, \bibinfo {author} {\bibfnamefont {M.~H.}\
  \bibnamefont {Devoret}}, \bibinfo {author} {\bibfnamefont {S.~M.}\
  \bibnamefont {Girvin}}, \ and\ \bibinfo {author} {\bibfnamefont {R.~J.}\
  \bibnamefont {Schoelkopf}},\ }\href {\doibase 10.1103/PhysRevA.76.042319}
  {\bibfield  {journal} {\bibinfo  {journal} {Phys. Rev. A}\ }\textbf {\bibinfo
  {volume} {76}},\ \bibinfo {pages} {042319} (\bibinfo {year}
  {2007})}\BibitemShut {NoStop}%
\bibitem [{\citenamefont {Schreier}\ \emph {et~al.}(2008)\citenamefont
  {Schreier}, \citenamefont {Houck}, \citenamefont {Koch}, \citenamefont
  {Schuster}, \citenamefont {Johnson}, \citenamefont {Chow}, \citenamefont
  {Gambetta}, \citenamefont {Majer}, \citenamefont {Frunzio}, \citenamefont
  {Devoret}, \citenamefont {Girvin},\ and\ \citenamefont
  {Schoelkopf}}]{schreier08a}%
  \BibitemOpen
  \bibfield  {author} {\bibinfo {author} {\bibfnamefont {J.~A.}\ \bibnamefont
  {Schreier}}, \bibinfo {author} {\bibfnamefont {A.~A.}\ \bibnamefont {Houck}},
  \bibinfo {author} {\bibfnamefont {J.}~\bibnamefont {Koch}}, \bibinfo {author}
  {\bibfnamefont {D.~I.}\ \bibnamefont {Schuster}}, \bibinfo {author}
  {\bibfnamefont {B.~R.}\ \bibnamefont {Johnson}}, \bibinfo {author}
  {\bibfnamefont {J.~M.}\ \bibnamefont {Chow}}, \bibinfo {author}
  {\bibfnamefont {J.~M.}\ \bibnamefont {Gambetta}}, \bibinfo {author}
  {\bibfnamefont {J.}~\bibnamefont {Majer}}, \bibinfo {author} {\bibfnamefont
  {L.}~\bibnamefont {Frunzio}}, \bibinfo {author} {\bibfnamefont {M.~H.}\
  \bibnamefont {Devoret}}, \bibinfo {author} {\bibfnamefont {S.~M.}\
  \bibnamefont {Girvin}}, \ and\ \bibinfo {author} {\bibfnamefont {R.~J.}\
  \bibnamefont {Schoelkopf}},\ }\href {\doibase 10.1103/PhysRevB.77.180502}
  {\bibfield  {journal} {\bibinfo  {journal} {Phys. Rev. B}\ }\textbf {\bibinfo
  {volume} {77}},\ \bibinfo {pages} {180502} (\bibinfo {year}
  {2008})}\BibitemShut {NoStop}%
\bibitem [{\citenamefont {Thompson}\ \emph {et~al.}(2013)\citenamefont
  {Thompson}, \citenamefont {Tiecke}, \citenamefont {de~Leon}, \citenamefont
  {Feist}, \citenamefont {Akimov}, \citenamefont {Gullans}, \citenamefont
  {Zibrov}, \citenamefont {Vuletic},\ and\ \citenamefont
  {Lukin}}]{thompson13a}%
  \BibitemOpen
  \bibfield  {author} {\bibinfo {author} {\bibfnamefont {J.~D.}\ \bibnamefont
  {Thompson}}, \bibinfo {author} {\bibfnamefont {T.~G.}\ \bibnamefont
  {Tiecke}}, \bibinfo {author} {\bibfnamefont {N.~P.}\ \bibnamefont {de~Leon}},
  \bibinfo {author} {\bibfnamefont {J.}~\bibnamefont {Feist}}, \bibinfo
  {author} {\bibfnamefont {A.~V.}\ \bibnamefont {Akimov}}, \bibinfo {author}
  {\bibfnamefont {M.}~\bibnamefont {Gullans}}, \bibinfo {author} {\bibfnamefont
  {A.~S.}\ \bibnamefont {Zibrov}}, \bibinfo {author} {\bibfnamefont
  {V.}~\bibnamefont {Vuletic}}, \ and\ \bibinfo {author} {\bibfnamefont
  {M.~D.}\ \bibnamefont {Lukin}},\ }\href {\doibase 10.1126/science.1237125}
  {\bibfield  {journal} {\bibinfo  {journal} {Science}\ }\textbf {\bibinfo
  {volume} {340}},\ \bibinfo {pages} {1202} (\bibinfo {year}
  {2013})}\BibitemShut {NoStop}%
\bibitem [{\citenamefont {Zhou}\ \emph {et~al.}(2008)\citenamefont {Zhou},
  \citenamefont {Gong}, \citenamefont {Liu}, \citenamefont {Sun},\ and\
  \citenamefont {Nori}}]{zhou08a}%
  \BibitemOpen
  \bibfield  {author} {\bibinfo {author} {\bibfnamefont {L.}~\bibnamefont
  {Zhou}}, \bibinfo {author} {\bibfnamefont {Z.~R.}\ \bibnamefont {Gong}},
  \bibinfo {author} {\bibfnamefont {Y.-x.}\ \bibnamefont {Liu}}, \bibinfo
  {author} {\bibfnamefont {C.~P.}\ \bibnamefont {Sun}}, \ and\ \bibinfo
  {author} {\bibfnamefont {F.}~\bibnamefont {Nori}},\ }\href {\doibase
  10.1103/PhysRevLett.101.100501} {\bibfield  {journal} {\bibinfo  {journal}
  {Phys. Rev. Lett.}\ }\textbf {\bibinfo {volume} {101}},\ \bibinfo {pages}
  {100501} (\bibinfo {year} {2008})}\BibitemShut {NoStop}%
\bibitem [{\citenamefont {Liao}\ \emph {et~al.}(2012)\citenamefont {Liao},
  \citenamefont {Cheung},\ and\ \citenamefont {Law}}]{liao12a}%
  \BibitemOpen
  \bibfield  {author} {\bibinfo {author} {\bibfnamefont {J.-Q.}\ \bibnamefont
  {Liao}}, \bibinfo {author} {\bibfnamefont {H.~K.}\ \bibnamefont {Cheung}}, \
  and\ \bibinfo {author} {\bibfnamefont {C.~K.}\ \bibnamefont {Law}},\ }\href
  {\doibase 10.1103/PhysRevA.85.025803} {\bibfield  {journal} {\bibinfo
  {journal} {Phys. Rev. A}\ }\textbf {\bibinfo {volume} {85}},\ \bibinfo
  {pages} {025803} (\bibinfo {year} {2012})}\BibitemShut {NoStop}%
\bibitem [{\citenamefont {Xu}\ and\ \citenamefont {Fan}()}]{xu15a}%
  \BibitemOpen
  \bibfield  {author} {\bibinfo {author} {\bibfnamefont {S.}~\bibnamefont
  {Xu}}\ and\ \bibinfo {author} {\bibfnamefont {S.}~\bibnamefont {Fan}},\
  }\href@noop {} {\bibinfo  {journal} {arXiv:1502.06049}\ }\BibitemShut
  {NoStop}%
\end{thebibliography}%

\newpage
\widetext
\begin{center}
\textbf{\large Supplemental Material:  \\Deterministic generation of photon pairs within the bad cavity limit}
\end{center}
\setcounter{equation}{0}
\setcounter{figure}{0}
\makeatletter

\renewcommand{\thefigure}{SM\arabic{figure}}
\renewcommand{\thesection}{SM\arabic{section}}  
\renewcommand{\theequation}{SM\arabic{equation}}  

This supplemental material is divided in three main sections. The first one, Sec.~\ref{Sec:ScatForm}, is devoted to the scattering formalism. After introducing the $S$-matrix (Sec.~\ref{Sec:Smat}), we proceed to connect the multi-time correlation functions of a weakly-driven system with the scattering wavefunctions (Sec.~\ref{Sec:Connection}). We then explain how to compute scattering amplitudes (Sec.~\ref{Sec:Samp}) and wavefunctions (Sec.~\ref{Sec:Swf}) in an operationally simple manner, and apply those methods to our cavity QED model (Sec.~\ref{Sec:cQEDscat}) proving that it can behave as a deterministic down-converter, and providing an analytical analysis of the photon-pair emission timescales. In Sec.~\ref{Sec:AlternativeForm} we provide an alternative formalism based on the master equation, which allows us to get insight into the properties of our cavity QED setup, including analytical expressions for its main observables. Sec.~\ref{Sec:cQEDimp} is the last one, and on it we show in detail how our 
proposed circuit QED architecture provides an implementation of the system Hamiltonian we are looking for.

\section{Scattering theory results}\label{Sec:ScatForm}

\subsection{Brief introduction to the $S$-matrix}\label{Sec:Smat}

The scattering formalism describes the asymptotic behaviour of free fields whose dynamics is governed by a Hamiltonian $H_B$, after having been scattered by some interacting Hamiltonian $H$. We use in particular the form of the Hamiltonian introduced in the main text
\begin{equation}\label{Hsup}
H=H_{S}+\underset{H_B}{\underbrace{\int\hspace{-1mm}dk\hspace{0.5mm}k(p_{k}^{\dagger }p_{k}+s_{k}^{\dagger }s_{k})}}+\underset{H_{S\hspace{-0.4mm}B}}{\underbrace{\int\hspace{-1mm}dk\left( 
\sqrt{\frac{\gamma _{p}}{2\pi }}p_{k}^{\dagger }a_{p}+\sqrt{\frac{\gamma _{s}%
}{2\pi }}s_{k}^{\dagger }a_{s}+\mathrm{H.c.}\right)}}.
\end{equation}
with a system Hamiltonian $H_S$ independent of the bath operators (the fields), which interact with the system through $H_{S\hspace{-0.2mm}B}$. Within the scattering formalism, all the asymptotic information can be extracted from the so-called $S$-matrix, defined by \cite{sakurai14a}
\begin{equation}
\label{eq:smatrix}
S=\lim\hspace{0.05mm}_{t_i\rightarrow-\infty}^{t_f\rightarrow+\infty}\hspace{1mm}e^{iH_Bt_f}e^{-iH(t_f-t_i)}e^{-iH_Bt_i}=\lim_{t\rightarrow\infty}\mathcal{T}\left[e^{-i\int_{-t}^{t}\tilde{H}(t')dt'}\right]\,,
\end{equation}
where we have moved to the interaction picture, where the Hamiltonian is transformed into
\begin{equation}
\tilde{H}(t)= e^{iH_Bt}(H_S+H_{S\hspace{-0.4mm}B}) e^{-iH_Bt} = H_S+\int\hspace{-1mm}dk\hspace{0.5mm}\left(\sqrt{\frac{\gamma_p}{2\pi}}p_{k}^{\dagger}e^{ikt}a_p+\sqrt{\frac{\gamma_s}{2\pi}}s_{k}^{\dagger}e^{ikt}a_s+\mathrm{H.c.}\right)\,.
\end{equation}
In addition, we have made use of the identity
\begin{equation}\label{PictureChange}
\mathcal{T}\left[e^{-i\int_{t_0}^t dt^\prime O(t^\prime)}\right]=e^{-i O_0 t}\mathcal{T}\left[\exp\left(-i\int_{t_0}^t dt^\prime e^{i O_0 t^\prime} O_1(t') e^{-iO_0 t^\prime}\right)\right] e^{i O_0 t_0}\,,
\end{equation}
valid for any operator $O(t)=O_0+O_1(t)$, where $\mathcal{T}$ (time-ordering operator) orders Heisenberg-picture operators $A(t)$ and $B(t)$ as
\begin{equation}
\mathcal{T}[A(t_1)B(t_2)]=\left\{\begin{array}{cc}
A(t_1)B(t_2) & \mathrm{if}\hspace{3mm}t_1>t_2
\\
B(t_2)A(t_1) & \mathrm{if}\hspace{3mm}t_1<t_2	
\end{array}\right.,
\end{equation}
allowing to write the Dyson series in the compact form
\begin{equation}\label{DysonSeries}
\mathcal{T}\left[e^{-i\int_{t_0}^t dt^\prime A(t')}\right]=1-i\int_{t_0}^t dt_1 A(t_1)-\int_{t_0}^t dt_1\int_{t_0}^{t_1} dt_2 A(t_1)A(t_2)+i\int_{t_0}^t dt_1\int_{t_0}^{t_1}dt_2\int_{t_0}^{t_2}dt_3 A(t_1)A(t_2)A(t_3)+...	
\end{equation}

\subsection{Connection between the correlation functions and scattering wavefunctions}\label{Sec:Connection}

In this section we connect the wavefunctions which are naturally defined within the scattering formalism with correlation functions of the driven system, which are the objects naturally measured in quantum-optical experiments. Consider then the $n$-th correlation function of the field leaking out of the system through the signal bath, which can be written as \cite{gardiner_book00a,carmichael_book02a}
\begin{equation}
G_{\mathrm{out}}^{(n)}(\tau _{1},\tau _{2},...,\tau _{n})=\bra{0}s_{\mathrm{out}}^{\dagger }(\tau _{1})...s_{\mathrm{out}}^{\dagger }(\tau_{n})s_{\mathrm{out}}(\tau _{n})...s_{\mathrm{out}}(\tau _{1})\ket{0}
\end{equation}
where $\ket{0}$ is the state with no excitations both in the system and the baths, we have assumed $\tau _{1}<...<\tau _{n}$, and defined the output field
\begin{equation}
s_{\mathrm{out}}(\tau )=(2\pi )^{-1/2}\lim_{t\rightarrow \infty }\int\hspace{-1mm}dk \hspace{1mm} s_{k}(t)e^{-ik(\tau -t)},
\end{equation}
with $s_{k}(t)=\mathcal{T}\left[ e^{i\int_{0}^{t}dt^{\prime }[H+H_D(t^{\prime })]}%
\right] s_{k}\mathcal{T}\left[ e^{-i\int_{0}^{t}dt^{\prime }[H+H_D(t^{\prime })]}%
\right] $ a bath operator in the Heisenberg picture, where we have included a driving term $H_{D}(t)=\Omega_{p}\left(e^{-ik_{0}t}a_{p}^{\dag }+\mathrm{H.c.}\right)$ to the total Hamiltonian. Note that the well-known input-output relation \cite{gardiner_book00a,carmichael_book02a} $s_{\mathrm{out}}(\tau )+s_{\mathrm{in}%
}(\tau )=\sqrt{\gamma _{s}}a_{s}(\tau )$, with
\begin{equation}
s_{\mathrm{in}}(\tau )=(2\pi )^{-1/2}\lim_{t\rightarrow 0}\int\hspace{-1mm}dk\hspace{1mm} s_{k}(t)e^{-ik(\tau -t)},
\end{equation}
together with $[a_{s}(\tau ),s_{\mathrm{in}}(\tau ^{\prime })]=0$ for $\tau
<\tau ^{\prime }$ (causality) and the fact that the signal bath is assumed
to be in vacuum at the origin of times, allows connecting $G_{\mathrm{out}%
}^{(n)}$ with the correlation functions of system operators that we defined
in the main text,
\begin{equation}
G_{\mathrm{out}}^{(n)}(\tau _{1},\tau _{2},...,\tau _{n})=\gamma _{s}^{n}%
\underset{G_{s}^{(n)}(\tau _{1},\tau _{2},...,\tau _{n})}{\underbrace{\left\langle a_{s}^{\dagger }(\tau _{1})...a_{s}^{\dagger }(\tau_{n})a_{s}(\tau _{n})...a_{s}(\tau _{1})\right\rangle }}.  \label{GoutG}
\end{equation}

The correlation function can be written as
\begin{equation}
G_{\mathrm{out}}^{(n)}(\tau _{1},\tau _{2},...,\tau _{n})=\lim_{t\rightarrow
\infty }\mathrm{tr}\left\{ s(x_{n})...s(x_{1})\mathcal{T}\left[
e^{i\int_{0}^{t}dt^{\prime }[H+H_{D}(t^{\prime })]}\right] \left\vert
0\right\rangle\hspace{-1mm} \left\langle 0\right\vert \mathcal{T}\left[
e^{-i\int_{0}^{t}dt^{\prime }[H+H_{D}(t^{\prime })]}\right] s^{\dagger
}(x_{1})...s^{\dagger }(x_{n})\right\} ,
\end{equation}%
where $s(x)=(2\pi )^{-1/2}\int\hspace{-1mm}dk\hspace{0.5mm}s_{k}e^{ikx}$ annihilates signal-field excitations in real space, and $x_{n}=t-\tau _{n}$. Let us now perform two unitary transformations inside the trace, which will allow us to easily get to the expression we are looking for. First, we
apply $U=\exp (-iH_{0}t)$ with $H_{0}=\int\hspace{-1mm}dk\hspace{0.5mm}kp_{k}^{\dagger }p_{k}$, which
turns the correlation function into%
\begin{equation}
G^{(n)}(\tau _{1},\tau _{2},...,\tau _{n})=\lim_{t\rightarrow \infty }%
\mathrm{tr}\left\{ s(x_{n})...s(x_{1})\mathcal{T}\left[ e^{-i\int_{0}^{t}%
\tilde{H}(t^{\prime })dt^{\prime }}\right] \left\vert 0\right\rangle\hspace{-1mm}\left\langle 0\right\vert \mathcal{T}\left[ e^{i\int_{0}^{t}\tilde{H}%
(t^{\prime })dt^{\prime }}\right] s^{\dagger }(x_{1})...s^{\dagger
}(x_{n})\right\} ,
\end{equation}
with
\begin{equation}
\tilde{H}(t)=U^{\dagger }(H+H_{D})U-H_{0}=H_{S}+\Omega _{p}\left(
e^{-ik_{0}t}a_{p}^{\dag }+\mathrm{H.c.}\right) +\int\hspace{-1mm}dk\hspace{0.5mm}ks_{k}^{\dagger
}s_{k}+\int\hspace{-1mm}dk\left( \sqrt{\frac{\gamma _{p}}{2\pi }}p_{k}^{\dagger
}e^{ikt}a_{p}+\sqrt{\frac{\gamma _{s}}{2\pi }}s_{k}^{\dagger }a_{s}+\mathrm{%
H.c.}\right) ,
\end{equation}
where we have made use of (\ref{PictureChange}). Let us now apply a displacement transformation $D$ on mode $k_{0}$ of the pump bath, defined through $D^{\dagger }p_{k}D=p_{k}+\alpha _{k_{0}}\delta
(k-k_{0})$ with $\alpha _{k_{0}}=\sqrt{2\pi /\gamma_p}\Omega _{p}$, which making use of (\ref{PictureChange}) again, turns the correlation function into
\begin{equation}
G_{\mathrm{out}}^{(n)}(\tau _{1},\tau _{2},...,\tau _{n})=\lim_{t\rightarrow
\infty }\mathrm{tr}\left\{ s(x_{n})...s(x_{1})\mathcal{T}\left[
e^{-i\int_{0}^{t}\bar{H}(t^{\prime })dt^{\prime }}\right] \left\vert \alpha
_{k_{0}}\right\rangle\hspace{-1mm} \left\langle \alpha _{k_{0}}\right\vert \mathcal{T}%
\left[ e^{i\int_{0}^{t}\bar{H}(t^{\prime })dt^{\prime }}\right] s^{\dagger
}(x_{1})...s^{\dagger }(x_{n})\right\} ,
\end{equation}%
where $\left\vert \alpha _{k_{0}}\right\rangle $ denotes the coherent state
of the pump mode $k_{0}$, and%
\begin{equation}
\bar{H}(t)=D^{\dagger }\tilde{H}D=H_{S}+\int\hspace{-1mm}dk\hspace{0.5mm}ks_{k}^{\dagger }s_{k}+\int\hspace{-1mm}dk\hspace{0.5mm}\left( \sqrt{\frac{\gamma _{p}}{2\pi }}p_{k}^{\dagger }e^{ikt}a_{p}+\sqrt{%
\frac{\gamma _{s}}{2\pi }}s_{k}^{\dagger }a_{s}+\mathrm{H.c.}\right) .
\end{equation}%
Using now the expansion of the coherent states in the Fock basis, and
undoing the $U$ transformation, we obtain%
\begin{equation}
G_{\mathrm{out}}^{(n)}(\tau _{1},\tau _{2},...,\tau _{n})=e^{-\alpha
_{k_{0}}^{2}}\sum_{lm=0}^{\infty }\frac{\alpha _{k_{0}}^{m+l}}{\sqrt{l!m!}}%
\lim_{t\rightarrow \infty }\left\langle 0\right\vert
p_{k_{0}}^{m}e^{iHt}s^{\dagger }(x_{1})...s^{\dagger
}(x_{n})s(x_{n})...s(x_{1})e^{-iHt}p_{k_{0}}^{\dagger l}\left\vert
0\right\rangle .
\end{equation}%
This expression is completely general, and its connection to scattering theory is
not entirely clear, since the $S$-matrix is defined as $S=\lim\hspace{0.05mm}_{t_i\rightarrow-\infty}^{t_f\rightarrow+\infty}\hspace{1mm}e^{iH_Bt_f}e^{-iH(t_f-t_i)}e^{-iH_Bt_i}$, while what appears on
it is $\lim_{t\rightarrow \infty }e^{\pm iHt}$. Let us then particularize the
expression to the case of interest for us, one in which the Hamiltonian $H$
can only generate signal photons pair by pair from pump photons. Under such
conditions, it is clear that the leading $\alpha _{k_{0}}$-order in the
previous expression is given by%
\begin{equation}
G_{\mathrm{out}}^{(n)}(\tau _{1},\tau _{2},...,\tau _{n})=\frac{\alpha
_{k_{0}}^{2\bar{m}}}{\bar{m}!}\lim_{t\rightarrow \infty }\left|\left\langle
0\right\vert s(x_{n})...s(x_{1})e^{-iHt}p_{k_{0}}^{\dagger \bar{m}%
}\left\vert 0\right\rangle \right|^{2},
\end{equation}%
where $\bar{m}$ is the minimum number of pump photons capable of generating $%
n$ signal photons, that is, $\bar{m}=\lceil n/2\rceil $, where $\lceil
x\rceil $ denotes the ceiling function (the smallest integer not smaller
than $n/2$). Now, it is also clear from the definition of the $S$-matrix that the matrix element in this expression and $\left\langle 0\right\vert
s(x_{n})...s(x_{1})Sp_{k_{0}}^{\dagger \bar{m}}\left\vert 0\right\rangle $
are equivalent up to a phase. Therefore, combining this expression with (\ref%
{GoutG}), we obtain the final result%
\begin{equation}
G_{s}^{(n)}(\tau _{1},\tau _{2},...,\tau _{n})\propto \Omega _{p}^{2\bar{m}%
}\lim_{t\rightarrow \infty }|\hspace{-1mm}\left\langle 0\right\vert
s(x_{n})...s(x_{1})Sp_{k_{0}}^{\dagger \bar{m}}\left\vert 0\right\rangle
\hspace{-1mm}|^{2}.
\end{equation}%
which particularized to to $n=2$ and $4$ coincide with the expressions
provided in the main text. This expression connects the multi-time
correlation functions of a weakly driven system with multi-photon wave
functions obtained from the scattering theory, what we use in the main text
to build a criterion characterizing when the system behaves as a proper single
two-photon source.

\subsection{Evaluation of scattering amplitudes}\label{Sec:Samp}

There exist severals methods to retrieve information from the S-matrix \cite{zhou08a,liao12a,caneva15a,shi15a,xu15a}, but here we will use the method introduced in Ref.~\cite{caneva15a,shi15a}, that we describe now. Within the scattering formalism one is naturally interested in how incoming pump-bath photons with well-defined momenta scatter into outgoing photons in the pump and signal baths (see Fig. \ref{fig:1} in the main text). The probabilities associated to such processes are encoded in the so-called \textit{scattering amplitudes}, which are nothing but the S-matrix elements connecting the desired Fock states of the baths. For illustration, we calculate in detail the simplest among these proceses, namely the so-called \textit{reflection amplitude}, which is the probability amplitude of a pump photon with momentum $k_i$ to transform into an outgoing pump photon with momentum $k_f$. This is given by
\begin{equation}
_p\langle k_f\vert S\vert k_i\rangle_p=\lim_{t\rightarrow\infty}\langle 0\vert p_{k_f} \mathcal{T}\left[e^{-i\int_{-t}^{t}\tilde{H}(t')dt'}\right] p_{k_i}^\dagger\vert 0\rangle\,,
\end{equation}
where we remind that $\ket{0}$ is the state with no excitations, that is, the common ground state of $H_B$ and $H_S$. This expression depends on both bath and system operators, and our approach is designed to transform it in such a way that it can be evaluated simply from matrix elements of system operators. In order to do so, we proceed as follows. First, note that the previous expression can be written with the use of functional derivatives as
\begin{equation}\label{ScattFunc}
_p\langle k_f\vert S\vert k_i\rangle_p=\lim_{t\rightarrow\infty}	\left.\frac{\delta}{\delta J_{k_f}^\ast}\frac{\delta}{\delta J_{k_i}}\langle 0\vert e^{\int\hspace{-1mm}dk J_k^\ast p_k} \mathcal{T}\left[e^{-i\int_{-t}^{t}\tilde{H}(t')dt'}\right] e^{\int\hspace{-1mm}dk J_k p_k^\dagger}\vert 0\rangle\right\vert_{\{J_k,J_k^\ast\}\rightarrow 0}
\end{equation}
where $\{J_k,J_k^\ast\}$ are treated as independent classical currents for each bath mode, and we remind that the functional derivative acts as
\begin{equation}
\frac{\delta}{\delta J_{k'}}J_k=\delta(k-k')\hspace{5mm}\Longrightarrow\hspace{5mm} \frac{\delta}{\delta J_{k'}}\exp\left(\int\hspace{-1mm}dk J_k A_k\right)=\exp\left(\int\hspace{-1mm}dk J_k A_k\right)A_{k'},
\end{equation}
valid for any operator $A_k$ (and similarly for $J_k^\ast$). Introducing the displacement operator for the pump bath modes
\begin{equation}
D(\{J_k\}) = \exp\left(\int\hspace{-1mm}dkJ_k p_k^\dagger\right)\exp\left(\int\hspace{-1mm}dkJ_k^\ast p_k\right)\exp\left(-\frac{1}{2}\int\hspace{-1mm}dk\vert J_k\vert^2\right) = \exp\left[\int\hspace{-1mm}dk\left(J_k p_k^\dagger - J_k^\ast p_k\right)\right],
\end{equation}
which transforms the Hamiltonian as $\bar{H}(t)=D^\dagger(\{J_k\}) \tilde{H}(t)D(\{J_k\}) = \tilde{H}(t)+H_D(t)$, that is, it adds a driving term
\begin{equation}
H_{D}(t)=\int\hspace{-1mm}dk\sqrt{\frac{\gamma_p}{2\pi}}\left(J_k^\ast e^{ikt}a_p+J_k e^{-ikt}a_p^\dagger\right)\,,
\end{equation}
we can rewrite Eq. (\ref{ScattFunc}) as
\begin{equation}
_p\langle k_f\vert S\vert k_i\rangle_p =\lim_{t\rightarrow\infty}\left.\frac{\delta}{\delta J_{k_f}^\ast}\frac{\delta}{\delta J_{k_i}}\langle 0\vert \mathcal{T}\left[e^{-i\int_{-t}^{t}dt_1\bar{H}(t_1)}\right]\vert 0\rangle e^{\int \hspace{-1mm}dk \vert J_k\vert^2}\right\vert_{\{J_k,J_k^\ast\}\rightarrow 0},
\end{equation}
and then come back from the interaction picture to get
\begin{equation}
_p\langle k_f\vert S\vert k_i\rangle_p = \lim_{t\rightarrow\infty}\left.\frac{\delta}{\delta J_{k_f}^\ast}\frac{\delta}{\delta J_{k_i}}\langle 0\vert \mathcal{T}\left[e^{-i\int_{-t}^{t}dt_1 [H+H_D(t_1)]}\right]\vert 0\rangle e^{\int \hspace{-1mm}dk \vert J_k\vert^2}\right\vert_{\{J_k,J_k^\ast\}\rightarrow 0}.
\end{equation}
Next we use the identity (\ref{PictureChange}), together the Dyson series (\ref{DysonSeries}) for the remaining time-ordered exponential, and the Taylor series of the normalization factor $e^{\int \hspace{-1mm}dk \vert J_k\vert^2}$, so that acting with the functional derivatives we easily obtain
\begin{equation}
_p\langle k_f\vert S\vert k_i\rangle_p = \delta(k_i-k_f)-\frac{\gamma_p}{2\pi} \int_{-\infty}^{\infty}dt_2\int_{-\infty}^{\infty}dt_1 \bra{0}\mathcal{T}[a_p(t_2)a_p^\dagger(t_1)]\ket{0} e^{ik_ft_2-ik_it_1},
\end{equation}
where given a Schr\"odinger-picture operator $A$, we have defined the Heisenberg-picture operator $A(t)=e^{iHt}Ae^{-iHt}$. This equation can be further simplified by assuming that $\ket{0}$ is annihilated by the system operators $a_j$, so that we can write
\begin{equation}
_p\langle k_f\vert S\vert k_i\rangle_p = \delta(k_i-k_f)-\frac{\gamma_p}{2\pi}\int_{-\infty}^{\infty}dt_2\int_{-\infty}^{t_2}dt_1 \langle 0\vert a_p e^{iH(t_1-t_2)} a_p^\dagger \vert 0\rangle e^{ik_ft_2-ik_it_1}.
\end{equation}
We can rewrite this expression in terms of system operators only as follows:
\begin{subequations}
\begin{eqnarray}
_p\langle k_f\vert S\vert k_i\rangle_p &=& \delta(k_i-k_f)-\frac{\gamma_p}{2\pi}\int_{-\infty}^{\infty}dt_2\int_{-\infty}^{t_2}dt_1 \mathrm{tr}\left\{a_p e^{-iH(t_2-t_1)} a_p^\dagger \vert 0\rangle\langle 0\vert e^{iH(t_2-t_1)}\right\} e^{ik_ft_2-ik_it_1}
\\
&=&\delta(k_i-k_f)-\frac{\gamma_p}{2\pi}\int_{-\infty}^{\infty}dt_{2}\int_{-\infty}^{t_2}dt_1 \mathrm{tr}_S \left\{ a_p e^{\mathcal{L}(t_2-t_1)} a_p^\dagger\vert 0\rangle_S\hspace{-0.1mm} \langle 0\vert\right\} e^{ik_ft_2-ik_it_1}\,, \label{TwoTimeCorrL}
\end{eqnarray}
\end{subequations}
where in the second equality we have performed the trace over the bath as $\mathrm{tr}_B\{e^{-iH\tau}A_Se^{iH\tau}\}=e^{\mathcal{L}\tau}[A_S]$, valid for any system operator $A_S$, by defining the Liouville superoperator
\begin{equation}\label{L}
\mathcal{L}(\cdot) =-i[H_S^\ast(\cdot)-(\cdot)H_S^{\ast\dagger}]+\gamma_p a_p(\cdot) a_p^\dagger+\gamma_s a_s(\cdot)a_s^\dagger,
\end{equation}
and the non-Hermitian system Hamiltonian
\begin{equation}
H_S^\ast=H_S-i\frac{\gamma_p}{2}a_p^\dagger a_p - i\frac{\gamma_s}{2} a_s^\dagger a_s\,.
\end{equation}%
$\ket{0}_S$ is the ground state of $H_S$. In order to get to the final operationally-friendly expression, let us denote by $\tilde{A}$ a system operator evolved with the non-Hermitian Hamiltonian $H_S^*$, that is, $\tilde{A}(t) = e^{iH_S^\ast t}Ae^{-iH_S^\ast t}$. The presence of $\ket{0}_S$ in (\ref{TwoTimeCorrL}), which we defined as ground state of $H_S$ and assumed to be annihilated by the system operators $a_j$, prevents any contribution from the jumps induced by the last two terms of (\ref{L}), so that (\ref{TwoTimeCorrL}) can be rewritten as
\begin{subequations}
\begin{eqnarray}
_p\langle k_f\vert S\vert k_i\rangle_p &=& \delta(k_i-k_f)-\frac{\gamma_p}{2\pi}\int_{-\infty}^{\infty}dt_2\int_{-\infty}^{t_2}dt_1 \bra{0}\tilde{a}_p(t_2) e^{ik_ft_2}\tilde{a}_p^\dagger(t_1)e^{-ik_it_1}\ket{0}
\\
&& \hspace{4cm}= \left(1+i\gamma_p\bra{0}a_p\frac{1}{H_S^*-k_i}a_p^\dagger\ket{0}\right) \delta(k_i-k_f)\,. \label{eqSM:ref}
\end{eqnarray}
\end{subequations}
Note how this expression allows evaluating scattering amplitudes by simply inverting the system operator $H_S^\ast-k_i$.

The scattering amplitude manipulated above is important as it determines the reflection coefficient of single photons sent to the system through the pump bath. In our case, there are another two important scattering amplitudes related to the emission of signal photons. The first one corresponds to the probability amplitude of transforming an input pump photon with momentum $k_i$ into two signal photons with momenta $\{q_1,q_2\}$, which following a similar approach as with the scattering amplitude above can be ultimately written as
\begin{subequations}\label{eqSM:2scat}
\begin{eqnarray}
_s\langle q_1,q_2\vert S\vert k_i\rangle_p &=& i\sqrt{\frac{\gamma_p}{2\pi}}\frac{\gamma_s}{2\pi}\int_{\mathbb{R}^3}\hspace{-1mm}dt_1dt_2dt_3 \hspace{1mm}e^{i(q_1t_3+q_2t_2-k_it_1)}\bra{0}\mathcal{T}\left[\tilde{a}_s(t_3)\tilde{a}_s(t_2)\tilde{a}_p^\dagger(t_1)\right]\ket{0} \label{TimeOrdered2}
\\
&=&-i\gamma_s\sqrt{\frac{\gamma_p}{2\pi}}\hspace{2mm}\langle 0\vert a_s\left[\frac{1}{H_S^*-q_1}+\frac{1}{H_S^*-q_2}\right]a_{s}\frac{1}{H_S^*-k_i} a_p^\dagger\vert 0\rangle \hspace{2mm}\delta(q_1+q_2-k_i) \,. \label{OpFriendly2}
\end{eqnarray}
\end{subequations}
Let us remark that the sum of two terms appearing inside the brackets appears because there are two different time-orderings which contribute, $\tilde{a}_s(t_3)\tilde{a}_s(t_2)\tilde{a}_p^\dagger(t_1)$ and $\tilde{a}_s(t_2)\tilde{a}_s(t_3)\tilde{a}_p^\dagger(t_1)$.

Similarly, the probability amplitude of transforming two input pump photons with momenta $\{k_1,k_2\}$ into four signal photons with momenta $\{q_1,q_2,q_3,q_4\}$ is given by a scattering amplitude which can written as
\begin{subequations}\label{eqSM:4scat}
\begin{eqnarray}	
_s\langle q_1,q_2,q_3,q_4\vert S\vert k_1,k_2\rangle_p &=& -\frac{\gamma_p}{2\pi}\left(\frac{\gamma_s}{2\pi}\right)^2\int_{\mathbb{R}^6}(\Pi_{i=1}^6 dt_i) e^{i(q_4t_6+iq_3t_5+q_2t_4+q_1t_3-k_2t_2-k_1t_1)}
\\
&&\hspace{3cm} \times \bra{0}\mathcal{T}\left[\tilde{a}_s(t_6)\tilde{a}_s(t_5)\tilde{a}_s(t_4)\tilde{a}_s(t_3)\tilde{a}_p^\dagger(t_2)\tilde{a}_{p}^\dagger(t_1)\right]\ket{0}, \nonumber
\end{eqnarray}
\end{subequations}
which can be trivially written in an operationally-friendly form similar to (\ref{eqSM:ref}) and (\ref{OpFriendly2}), but which is too lengthy to be written here since there many time-orderings which give nonzero contribution.

\subsection{Evaluation of the two- and four-photon wavefunctions}\label{Sec:Swf}

The last two scattering amplitudes that we have given above,  in Eqs. (\ref{eqSM:2scat}) and (\ref{eqSM:4scat}), are interesting because their Fourier transform provides the two- and four-photon wavefunctions introduced in the main text. Let us here provide closed expressions for these wavefunctions which can be evaluated directly from system operators.

In the case of the two-photon wavefunction, it is simple but lengthy by using (\ref{TimeOrdered2}) to obtain
\begin{equation}
\hspace{-0.mm}\Psi_{2\mathrm{ph}}(x_1,x_2) \hspace{-0.5mm}= \hspace{-0.5mm}\frac{1}{2\pi}\hspace{-1.4mm}\int\hspace{-1mm}dq_1 \hspace{-2mm}\int\hspace{-1mm}dq_2 \hspace{-0.5mm}\left._s\hspace{-0.2mm}\langle q_1,\hspace{-0.2mm}q_2\vert S\vert k_i\rangle_p\right.\hspace{-1mm}e^{i(q_1 \hspace{-0.2mm}x_1\hspace{-0.3mm}+q_2\hspace{-0.2mm}x_2)}\hspace{-1mm}=\hspace{-1mm}\sqrt{\frac{\gamma_p}{2\pi}}\gamma_s e^{ik_i\mathrm{max}\{x_1\hspace{-0.3mm},x_2\}}\hspace{-0.5mm}\langle 0\vert a_s e^{-iH_S^*\vert x_1-x_2\vert}a_s\frac{1}{H_S^*-k_i}a_p^\dagger\vert 0\rangle.\hspace{-0.9mm}\label{eqSM:2ph}
\end{equation}
Note that, since the eigenvalues of $H_{S}^*$ always have negative imaginary part, $\Psi_\mathrm{2ph}(x_1,x_2)\rightarrow 0$ when $\vert x_1-x_2\vert\rightarrow\infty$, showing that signal photons are emitted with a finite delay. From a practical point of view, this also means that this wavefunction has no independent scattering contribution. Therefore, in order to normalize it, one, e.g.,  can divide by its maximal value in $r$ (or equivalently $\tau$) as we did in Fig. \ref{fig:2}(b) of the main text.

The four-photon wavefunction
\begin{equation}
\Psi_{4\mathrm{ph}}(x_1,x_2,x_3,x_4) = \frac{1}{(2\pi)^2}\int\hspace{-1mm}dq_1 \int\hspace{-1mm}dq_2 \int\hspace{-1mm}dq_3 \int\hspace{-1mm}dq_4 \left._s\langle q_1,q_2,q_3,q_4\vert S\vert k_1,k_2\rangle_p\right. e^{i(q_1 x_1+q_2x_2+q_3x_3+q_4x_4)}
\end{equation}
can also be easily found, but is much more elaborate (and has a much lengthier final expression) in the general case. However, there are three relevant limits in which it is greatly simplified (we further assume $x_1>x_2>x_3>x_4$). First the limits $x_1-x_2\rightarrow\infty$ or $x_3-x_4\rightarrow\infty$, in which it vanishes identically, $\Psi_{4\mathrm{ph}}\rightarrow 0$, showing that the distance between the photons within the same pair is always finite, in agreement with what we found from the two-photon wavefunction. Second, the limit $x_2-x_3\rightarrow\infty$, which gives information about how the photons within the pairs behave when the system emits well-spaced pairs, and can be written as
\begin{equation}\label{eqSM:4ph}
\Psi_\mathrm{4ph}(x_1,x_2,x_3,x_4) = \left.\Psi_\mathrm{2ph}(x_1,x_2)\right\vert_{k_i=k_1}\left.\Psi_\mathrm{2ph}(x_3,x_4)\right\vert_{k_i=k_2} + \left.\Psi_\mathrm{2ph}(x_1,x_2)\right\vert_{k_i=k_2}\left.\Psi_\mathrm{2ph}(x_3,x_4)\right\vert_{k_i=k_1},
\end{equation}
which shows only independent scattering contributions where two incoming pump photons are scattered by the system independently into signal pairs described by the two-photon wavefunction (\ref{eqSM:2ph}). Thus, in this case we can normalize the general four-photon wavefunction to this independent scattering contribution, allowing us to compare directly with the normalized $g_{s,2}^{(2)}(\tau)$ shown in Fig. \ref{fig:2}(c) of the main text.

Finally, we consider the limit $x_1-x_2\rightarrow 0$ and $x_3-x_4\rightarrow 0$, which assumes the photons within the pairs to overlap perfectly and thus gives information about the relative distance $R=x_2-x_3$ between the the two-photon wavepackets (describing then the bunching or antibunching between the photon pairs). We get $\Psi_\mathrm{4ph}(x_1,x_2,x_3,x_4)=\psi_\mathrm{4ph}(R;k_1,k_2)+\psi_\mathrm{4ph}(R;k_2,k_1)$, with
\begin{eqnarray}\label{4phWF}
\psi_\mathrm{4ph}(R,k_1,k_2) &=& -\frac{\gamma_p}{2\pi}\gamma_s^2 e^{i(k_1+k_2)x_1}\left[\bra{0}a_s^2 e^{-iH_S^*R}a_s^2\frac{1}{k_1+k_2-H_S^*}a_p^\dagger\frac{1}{H_S^*-k_1}a_p^\dagger\ket{0}\right.
\\
&&\hspace{4cm}\left.+\bra{0}a_s^2 \frac{e^{-iH_S^*R}-e^{-ik_2R}}{H_S^*-k_2}a_p^\dagger\ket{0}
\bra{0} a_s^2\frac{1}{H_S^*-k_1}a_p^\dagger\ket{0}\right].  \nonumber
\end{eqnarray}

\subsection{Application to our system Hamiltonian: condition for deterministic down-conversion and analysis of timescales}\label{Sec:cQEDscat}

Let us now particularize the previous general results to our cavity QED system Hamiltonian, Eq. (5) in the main text, which we reproduce here for completeness
\begin{equation}\label{HsSM}
H_S = g_p\,  a_p^\dagger\ketbra{g}{e} + \Omega_s\ketbra{m_2}{m_1} + g_s\, a_s^\dagger \left(\ketbra{m_2}{e}+\ketbra{g}{m_1}\right)+\mathrm{H.c.}\,,	
\end{equation}
where the states $\{\ket{g},\ket{m_1},\ket{m_2},\ket{e}\}$ form a four-level system and $a_j$ are bosonic annihilation operators associated to two cavity modes. It is important to note that the operator $C= 2a_p^\dagger a_p + 2\ketbra{e}{e} + a_s^\dagger a_s + \ketbra{m_1}{m_1} + \ketbra{m_2}{m_2}$ commutes with the Hamiltonian, and hence, $H_S^*$ does not connect subspaces with different eigenvalue $c$ of $C$. In particular, let us define the basis $\{\ket{n}_p\otimes\ket{l}_s\otimes\ket{r}=\ket{n,l,r}\}_{n,l=0,1,2,...}^{r=g,m_1,m_2,e}$, where $\vert n\rangle_j$ is a Fock state for cavity mode $j$. Then the representation of $H_S^\ast$ will have a box structure, each box corresponding to a subspace with well defined eigenvalue, e.g., $c=0$ spanned by $\{\vert 0,0,g\rangle\}$, $c=1$ spanned by $\{\ket{0,1,g},\ket{0,0,m_1},\ket{0,0,m_2}\}$, $c=2$ spanned by $\{\ket{1,0,g},\ket{0,2,g},\ket{0,0,e},\ket{0,1,m_1},\ket{0,1,m_2}\}$, $c=3$ spanned by $\{\ket{1,1,g},\ket{0,3,g},\ket{1,0,m_1},\ket{1,0,m_2},\ket{0,1,e},
\ket{0,2,m_
1},\ket{0,2,m_2}\}$, or $c=4$ spanned by $\{\ket{2,0,g},\ket{1,2,g},\ket{0,4,g},\ket{0,2,e},\ket{1,0,e},\ket{0,3,m_1},\ket{1,1,m_1},\ket{1,1,m_2},\ket{0,3,m_2}\}$. In fact, these are all the subspaces that we need to consider in order to evaluate the main scattering amplitudes and wavefunctions introduced above.

For example, consider the reflection amplitude of an incoming pump photon, Eq. (\ref{eqSM:ref}); it is clear that this can be evaluated by considering the $c=2$ subspace of $H_S^\ast-i\gamma^\ast\ketbra{e}{e}/2$ (where we consider also spontaneous emission of $\ket{e}$ to $\ket{g}$), which can be written as:
\begin{equation}\label{Hs2star}
H_{S2}^\ast = \left(\begin{array}{ccccc}
-i\gamma_p/2 & 0 & g_p & 0 & 0 \\ 
0 & -i\gamma_s & 0 & \sqrt{2}g_s & 0 \\ 
g_p & 0 & -i\gamma^*/2 & 0 & g_s \\ 
0 & \sqrt{2}g_s & 0 & -i\gamma_s/2 & \Omega_s^\ast \\ 
0 & 0 & g_s & \Omega_s & -i\gamma_s/2
\end{array}\right).
\end{equation}
The reflection amplitude can be obtained then as
\begin{equation}
_p\langle k_f\vert S\vert k_i\rangle_p = \left[1+i\gamma_p(H_{S2}^*-k_iI_{5\times 5})^{-1}_{11}\right]\delta(k_i-k_f),
\end{equation}
where $I_{5\times 5}$ is the $5\times 5$ identity matrix and $(H_{S2}^*-k_iI_{5\times 5})^{-1}_{11}$ refers to element 11 of the inverse of the matrix $H_{S2}^*-k_iI_{5\times 5}$, which can be easily found with the help of any symbolic program. Let us define the reflection coefficient $\mathcal{R}_p(k_i)=\vert 1+i\gamma_p(H_{S2}^*-k_iI_{5\times 5})^{-1}_{11}\vert^2$. Even though its full expression is too lengthy to be shown here, we have checked by exhaustive numerical inspection that within the bad-cavity limit it can be very well approximated by the Lorentzian
\begin{equation}
\mathcal{R}_p(k_i) = \left\vert 1-\frac{2 \Gamma_p}{\Gamma_p+\gamma^*+\Gamma_s(\Omega_s)-2ik_i}\right\vert^2,
\end{equation}
where we have introduced the Purcell rates $\Gamma_p=4g_p^2/\gamma_p$ and $\Gamma_s(\Omega_s)=4g_s^2\gamma_s/(\gamma_s^2+4\Omega_s^2)$, and we have assumed $\gamma^*\ll \Gamma_p,\Gamma_s$. We see that deterministic down-conversion can be obtained only at resonance $k_i=0$ and by demanding $_p\langle k_f\vert S\vert k_i = 0\rangle_p = 0\hspace{2mm}\forall k_f$, leading to an optimal control drive $\Omega_s$ given by
\begin{equation}\label{Omega2phSup}
\Omega_\mathrm{2ph}^2 = \frac{\gamma_s^2}{4}\frac{\Gamma_s(0)-(\Gamma_p-\gamma^*)}{\Gamma_p-\gamma^\ast}\,,
\end{equation}
expression that coincides with the one given in the main text in limit of negligible spontaneous emission ($\gamma^\ast \rightarrow 0$), Eq. (7), generalizing it to the case where spontaneous emission is present. Note that even in the presence of spontaneous emission it is still possible to obtain deterministic down-conversion, provided that certain conditions are satisfied. In particular, defining the cooperativites $C_j=g_j^2/\gamma_j\gamma^*$, we see that assuming $C_p>1/4$ then the condition $C_s>C_p-1/4$ is required. 

Proceeding in a similar way but making use of all the subspaces up to $c=4$, we can find analytical expressions for the two- and four-photon wavefunctions. These analytical expressions will allow us to see how the different timescales of photon-pair emission depend on the system parameters. For example, the two-photon wave function (\ref{eqSM:2ph}), it is easy to show that for resonant injection ($k_i=0$), and within the bad-cavity limit ($g_j\ll\gamma_j$) together with a strong control drive ($\Omega_s\gg g_s^2/\gamma_s$), we can write the corresponding probability as
\begin{equation}
\left\vert\Psi_\mathrm{2ph}(x_1,x_2)\right\vert^2 \hspace{-0.8mm}\propto\hspace{-0.8mm} \left\vert 2\Omega_s\exp\hspace{-1mm}\left[-\left(\gamma_s-\Gamma_s(\Omega_s)\right) r/2\right]\hspace{-1mm}-\hspace{-0.7mm}\gamma_s\exp\hspace{-1mm}\left(-\Gamma_s(\Omega_s)r/2\right)\sin\hspace{-1mm}\left[\Omega_s\hspace{-1mm}\left(1+\frac{\Gamma_s(\Omega_s)}{2\gamma_s}\right)r\right]\right\vert^2\hspace{-1.5mm},
\end{equation}
where $r=\vert x_1-x_2\vert$. It is easy to find by numerical inspection that the value of $r$ that maximizes this expression scales as $\Omega_s^{-1}$ for the regime of interest, i.e., $\Omega_s\gg \gamma_s (\gg \Gamma_s(0))$. This quantity is directly related to the timescale separation between the photons within the pair, denoted by $\tau_\mathrm{in}$ in the main text, which we therefore find to scale as $\tau_\mathrm{in}\propto\Omega_s^{-1}$. Hence, the photons within the pair can be made to overlap very well simply by working with large $\Omega_s$. On the other hand, under these circumstances the width of the two-photon wavefunction coincides with the intrinsic width of the emitted photons, denoted by $\tau_B$ in the main text, which for $\Omega_s\gg\gamma_s$ is simply given by the width of the first term in the expression above, that is, $\tau_B\approx \gamma_s^{-1}$. Thus, within the bad-cavity limit the $\Omega_s\gg 
\gamma_s$ condition ensures that the signal photon-pairs are emitted within the same temporal 
pulse (or spatial wavepacket within the scattering formalism) since $\tau_\mathrm{in}\ll\tau_B$.

In order to study the separation or bunching between the photon pairs, what we called $\tau_A$ in the main text, we analyze the four-photon wavefunction $\Psi_\mathrm{4ph}(x_1,x_2,x_3,x_4)$ under the assumption that we satisfy the condition $\tau_\mathrm{in}\ll\tau_B$ as explained above. Under such conditions, we can fix $x_{1}-x_{2}\approx 0$ and $x_{3}-x_{4}\approx 0$, and use then expression (\ref{4phWF}) to study the four-photon wavefuntion at resonance ($k_1=k_2=0$). To get the timescale $\tau_A$ of antibunching between the pairs, we then need to study the dependence of (\ref{4phWF}) with the relative coordinate $R=x_2-x_3$ between the wavepackets. Even though it is easy to find an analytic expression for it, it is too lengthy and nothing is gained from reproducing it here. However, in the bad-cavity limit it is possible to show that to second order in $g_j/\gamma_j$ we get $\vert\psi_\mathrm{4ph}(R,k_1,k_2)\vert^2\propto |1-e^{-R/2\tau_A}|^2$, with
\begin{equation}
\tau_A^{-1}=\Gamma_s(\Omega_s)+\Gamma_p\,,\label{eqSM:anti}
\end{equation}
in the absence of spontaneous emission. This explains why in Fig. \ref{fig:2}(c), the curves with $\Omega_s\gg\gamma_s$ or $\Omega_s\ll\gamma_s$ collapse to a single antibunched curve with respective growth rates $\Gamma_p$ or $\Gamma_p+\Gamma_{s}(0)$. When spontaneous emission is included this timescale must be modified. In particular, when $\vert \gamma_j-\gamma^*\vert\gg g_j$ is satisfied, we get
\begin{equation}
\tau_A^{-1}\approx\frac{2\left( \gamma _{s}-\gamma^*\right) g_{s}^{2}}{%
4\Omega_s^{2}+\left(\gamma^*-\gamma _{s}\right) ^{2}}+\frac{%
2g_{p}^{2}}{\gamma _{p}-\gamma^*}+\frac{\gamma^*}{2}\,,
\end{equation}
which converges to the value provided in the main text when $\gamma^*\ll \gamma_j$.

\section{Effective cavity QED model: efficiencies and timescales.}\label{Sec:AlternativeForm}
 
\begin{figure}
\begin{center}
\includegraphics[width=0.7\linewidth]{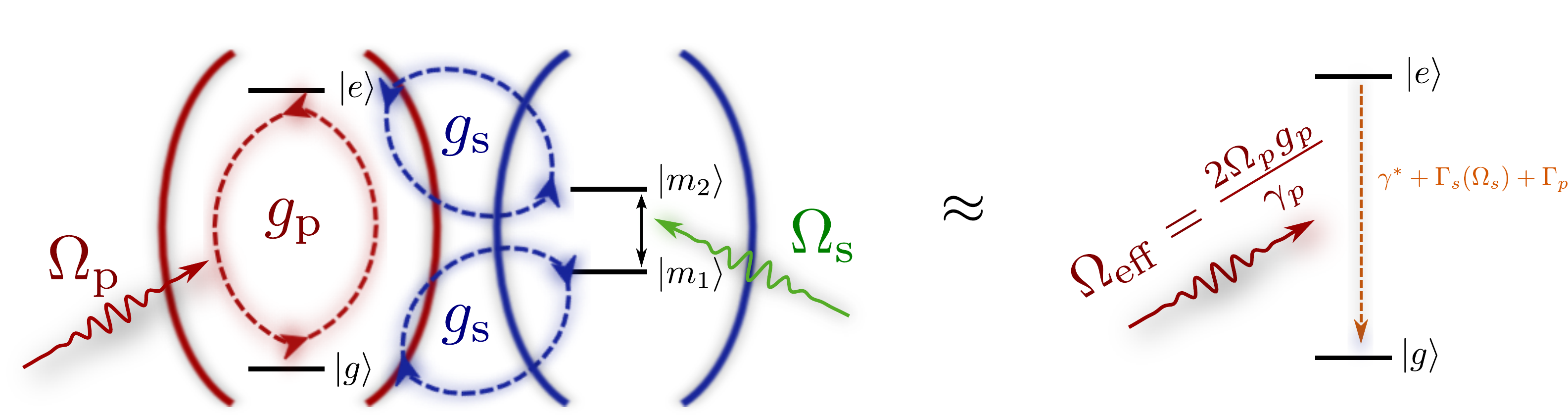}
\end{center}
\caption{Effective two-level system resulting from the adiabatic elimination of all the degrees of freedom except for the ground and excited states. The effective system has an effective decay rate $\gamma_\text{eff}=\Gamma_p+\Gamma_s(\Omega_s)+\gamma^*$ and effective coherent driving $\Omega_\text{eff}=2\Omega_p g_p/\gamma_p$.}
\label{figSM:3}
\end{figure}

This section aims at providing an intuitive understanding of our system by reducing it to a simpler one that allows explaining the underlying physics in an easier manner. In the previous section, we have shown how the interesting regime for the emission of photon pairs is the bad-cavity limit $g_j\ll\gamma_j$, together with $\Omega_s\gg\gamma_s$ such that photons are bunched. Under these conditions, the populations in the signal cavity and the intermediate levels $\{\ket{m_1},\ket{m_2}\}$ are very small, while the population in the pump cavity is basically given by the driving laser, that is, $\langle a_p\rangle\approx 2\Omega_p/\gamma_p$. This allows for an adiabatic elimination of these degrees of freedom, such that the dynamics is captured by an effective two-level model defined by the states $\{\ket{g},\ket{e}\}$, see Fig. \ref{figSM:3}. In this reduced model, the coherent amplitude of the pump cavity acts as a driving term $\Omega_\text{eff}(\ketbra{g}{e}+\ketbra{e}{g})$ with $\Omega_\text{eff}=2\Omega_
p g_p/\gamma_p$. In addition, there is an effective decay rate given by $\gamma_\text{eff}=\Gamma_p+\Gamma_s(\Omega_s)+\gamma^*$, which is easily found by applying perturbation theory in the $c=2$ subspace Hamiltonian (\ref{Hs2star}). Note that this effective rate coincides with the reloading rate (\ref{eqSM:anti}) that we found in the previous section.

The advantage of using this simplified model is two-fold: on the one hand it gives an intuitive and simplified picture of the dynamics of the system; moreover, due to the simplicity of the effective model, we can obtain analytical formulas from which reading the scaling of the relevant magnitudes, e.g, for the population of the signal cavity, $n_s$, which is related to the efficiency of two-photon emission. In order to obtain to obtain $n_s$, we will first obtain the population of the effective two level system $n_e=\text{tr}\{\rho\ketbra{e}{e}]$. In the simplified model $n_e$ corresponds to the excited state population of a driven-dissipative two-level system, which is given by
\begin{equation}
\label{eqSM:ne}
n_e=\frac{4\Omega_{\mathrm{eff}}^2}{\gamma_\text{eff}^2+8\Omega_\mathrm{eff}^2}=\frac{4g_p^2\Omega_{p}^2}{[\Gamma_p+\Gamma_s(\Omega_s)+\gamma^*]^2\gamma_p^2+8\Omega_{p}^2 g_p^2}\,,
\end{equation}
The number $N$ of transitions from $\ket{e}$ to $\ket{g}$ per unit time going through the signal cavity is given by $N = \Gamma_s(\Omega_s)n_e$, whereas the number of photons $N_\mathrm{s}$ emitted per unit time from the signal cavity is given by $N_\mathrm{s}=\gammas n_s$. On the other hand, any time that one of the $\ket{e}\rightarrow\ket{g}$ transitions takes place through the signal cavity, two photons are emitted, and hence $N_\mathrm{s} = 2 N$, leading to
\begin{equation}
n_s=2\frac{\Gamma_s(\Omega_s)}{\gammas}n_e
\label{eq:ns}
\end{equation}
Maximizing this expression with respect to the control drive $\Omega_s$, we obtain its optimal value. Interestingly, within the bad-cavity limit, this value coincides with the one obtained by demanding deterministic down-conversion in the scattering formalism, Eq. (\ref{Omega2phSup}).

\section{Details about the circuit QED implementation}\label{Sec:cQEDimp}

\begin{figure}[tb]
\begin{center}
\includegraphics[width=0.99\linewidth]{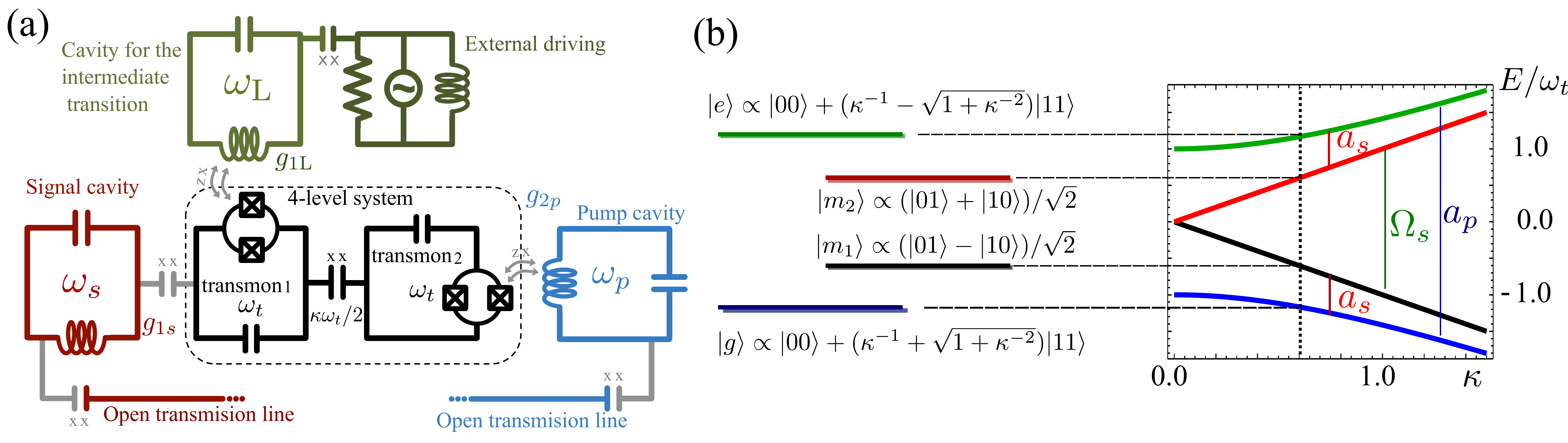}
\end{center}
\caption{(a) Circuit QED implementation: two coupled transmon qubits provide the desired four-level structure (b), while three LC resonators provide the single-mode cavities playing the role of signal, pump, and the classical driving for the intermediate transition. The baths are implemented with open transmission lines.}
\label{figSM:4}
\end{figure}

In this section, we show how the circuit QED architecture introduced in the main text provides an implementation of the nonlinear cavity QED setup that we have proposed as a photon-pair source.
The four-level structure is obtained from two identical qubits coupled through an $xx$ interaction, two capacitively-coupled transmons \cite{koch07a,schreier08a} in Fig. \ref{figSM:4}(a), described by the Hamiltonian
\begin{equation}
H_{2\mathrm{qb}}=\frac{\omega_t}{2}\left(\sigma_z^{(1)}+\sigma_z^{(2)} + \kappa\sigma_x^{(1)}\sigma_x^{(2)}\right)\,,
\end{equation}
where $\sigma^{(j)}_k$ is the $k$'th Pauli matrix associated to qubit $j$, and we will denote by $\ket{0}$ and $\ket{1}$, respectively, the eigenvectors of $\sigma_z$ with $-1$ and $+1$ eigenvalue (which for a transmon operated at the sweet-spot of the charge bias correspond, respectively, to the sum and the difference of the two lowest eigenstates of the circuit's charge operator). The spectrum of this Hamiltonian is plotted in Fig. \ref{figSM:4}(b), and it consists of the energies $E_{2,1}=\pm\omega_t\kappa$ corresponding to the symmetric and antisymmetric single-excited states $\ket{m_{2,1}}\propto\ket{10}\pm\ket{01}$, and the energies $E_{e,g}=\pm\omega_t\sqrt{1+\kappa^2}$ corresponding to the combinations $\ket{e,g}\propto (\kappa^{-1}\mp\sqrt{1+\kappa^{-2}})\ket{11}+\ket{00}$. Remarkably, the spectrum contains four levels with exactly the energy spacings that we require in our cavity QED proposal.

The next step consist on coupling the qubits to single-mode cavities in such a way that we reproduce the couplings that appear in the Hamiltonian $H_S$, Eq. (\ref{HsSM}). There are several ways of accomplishing this. For example, we consider three LC resonators with characteristic frequencies $\omega_p$ (pump), $\omega_\mathrm{L}$ (`laser'), and $\omega_s$ (signal). The pump and laser resonators are coupled to qubits 2 and 1, respectively, through an $xz$ interaction represented as an inductive coupling in Fig. \ref{figSM:4}(a); the signal cavity, on the other hand, is capacitively coupled through an $xx$ interaction to the first qubit. The corresponding Hamiltonians read
\begin{subequations}\label{CouplingHcQED}
\begin{eqnarray}
H_{1\mathrm{L}} &=& g_{1\mathrm{L}}(a_\mathrm{L}+a_\mathrm{L}^\dagger)\sigma_z^{(1)} \approx g_{1\mathrm{L}}(\alpha e^{-i\omega_\mathrm{L}t}+\alpha^\ast e^{i\omega_\mathrm{L}})\sigma_z^{(1)}\,,
\\
H_{2p} &=& g_{2p}(a_p+a_p^\dagger)\sigma_z^{(2)}\,,
\\
H_{1s} &=& g_{1s}(a_s+a_s^\dagger)\sigma_x^{(1)}\,,
\end{eqnarray}
\end{subequations}
where $a_j$ is the annihilation operator of the corresponding cavity, and in the first Hamiltonian we have assumed that the laser cavity is strongly driven, so that $a_\mathrm{L}$ can be replaced by its expectation value $\langle a_\mathrm{L}\rangle=\alpha e^{-i\omega_\mathrm{L}t}$.

We are going to show that, under certain conditions, these Hamiltonian terms provide the ones that we are looking for when written in the eigenbasis $\{\ket{g},\ket{m_1},\ket{m_2},\ket{e}\}$ of the coupled-qubit system. To this aim, let us first remark that in such basis the single-qubit operators are written as
\begin{subequations}
\begin{eqnarray}
\sigma_z^{(1)} &=& \ketbra{e}{e}-\ketbra{g}{g}-\kappa\ketbra{g}{e}+\ketbra{m_1}{m_2}+\mathrm{H.c.},
\\
\sigma_z^{(2)} &=& \ketbra{e}{e}-\ketbra{g}{g}-\kappa\ketbra{g}{e}-\ketbra{m_1}{m_2}+\mathrm{H.c.},
\\
\sqrt{2}\sigma_x^{(1)} &=& \ketbra{m_2}{e}-\ketbra{m_1}{e}+\ketbra{g}{m_2}+\ketbra{g}{m_1}+\mathrm{H.c.},
\end{eqnarray}
\end{subequations}
where, for the sake of simplifying the expressions, we have assumed $\kappa\ll 2$, though this is not really demanded for our system to work  (the only requirement being that $\kappa$ is not much bigger than 1 as this would make $\ket{m_{1,2}}$ and $\ket{g,e}$ degenerate). Hence, in an interaction picture defined with respect to the free Hamiltonian $H_\mathrm{2qb}+\omega_s a_s^\dagger a_s + \omega_p a_p^\dagger a_p$, the coupling Hamiltonians (\ref{CouplingHcQED}) can be written as
\begin{subequations}
\begin{eqnarray}
H_{1\mathrm{L}}(t) &=& g_{1\mathrm{L}}\alpha^* e^{i\omega_\mathrm{L}t}\left(\ketbra{e}{e}-\ketbra{g}{g}-\kappa\ketbra{g}{e}e^{-2i\omega_t\sqrt{1+\kappa^2}t}+\ketbra{m_1}{m_2}e^{-2i\omega_t\kappa t}+\mathrm{H.c.}\right)+\mathrm{H.c.},
\\
H_{2p}(t) &=& g_{2p}a_p^\dagger e^{i\omega_p t}\left(\ketbra{e}{e}-\ketbra{g}{g}-\kappa\ketbra{g}{e} e^{-2i\omega_t\sqrt{1+\kappa^2}t}-\ketbra{m_1}{m_2}e^{-2i\omega_t\kappa t}+\mathrm{H.c.}\right)+\mathrm{H.c.},
\\
H_{1s}(t)&=&\frac{g_{1s}}{\sqrt{2}} a_s^\dagger e^{i\omega_s t}\hspace{-1mm}\left[(\ketbra{m_2}{e}+\ketbra{g}{m_1})e^{i\omega_t(\kappa-\sqrt{1+\kappa^2})t}+(\ketbra{g}{m_2}-\ketbra{m_1}{e})e^{-i\omega_t(\kappa+\sqrt{1+\kappa^2})t}\hspace{-0.5mm}+\mathrm{H.c.}\right]+\mathrm{H.c.}.\hspace{1.4cm}
\end{eqnarray}
\end{subequations}
We therefore see that choosing the frequencies of the LC resonators as $\omega_\mathrm{L}=2\omega_t\kappa$, $\omega_p=2\omega_t\sqrt{1+\kappa^2}$, $\omega_s=\omega_t(\sqrt{1+\kappa^2}-\kappa)$, we make resonant the couplings $\{\alpha\ketbra{m_2}{m_1},a_p^\dagger\ketbra{g}{e},a_s^\dagger(\ketbra{m_2}{e}+\ketbra{g}{m_1})\}$ present in the target Hamiltonian (\ref{HsSM}), while the rest of the terms are suppressed within a rotating-wave approximation as long as $\{g_{1\mathrm{L}}\alpha,g_{2p}\kappa,g_{1s}/\sqrt{2}\}\ll 2\omega_t\kappa$. Under such conditions, we recover the desired Hamiltonian with the identifications $\Omega_s=g_{1\mathrm{L}}\alpha$, $g_p=-g_{2p}\kappa$, and $g_s=g_{1s}/\sqrt{2}$.

\end{document}